\documentclass[twocolumn,rsc,amsmath,showpacs]{revtex4}

\usepackage{amssymb}
\usepackage{amsmath}
\usepackage{color}

\usepackage{graphicx}

\begin{document}

\title{A microscopic model for colloidal 
gels with directional effective interactions:\\
Network induced glassy dynamics.}
\author{Emanuela Del Gado$^{1}$, 
and Walter Kob$^{2}$}
\affiliation{$^{1}$ETH Z\"urich, Department of Materials, Polymer Physics 
,
CH-8093 Z\"urich, Switzerland\\
$^{2}$ Laboratoire des Collo\"\i des, Verres et Nanomat\'eriaux, Universit\'e
Montpellier 2 and CNRS,
34095 Montpellier, France}
\date{\today}

\begin{abstract}
By means of molecular dynamics, we study the structure and the dynamics of a 
microscopic model for colloidal gels at low volume fractions. 
The presence of directional interactions leads to the formation 
of a persistent interconnected network at temperatures where phase 
separation does not occur. 
We find that large scale spatial correlations strongly depend on the volume 
fraction and characterize the formation of the persistent network. 
We observe a pre-peak in the static structure factor and relate it to the network 
structure. The slow dynamics at gelation is characterized by the coexistence
of fast collective motion of the mobile parts of the network structure (chains)
with large scale rearrangements producing stretched exponential relaxations.
We show that, once the network is sufficiently persistent, it
induces slow, cooperative processes related to the network nodes. 
We suggest that this peculiar glassy dynamics is a hallmark of the 
physics of colloidal gels at low volume fractions.
\end{abstract}


\maketitle
\section{Introduction}
\label{intro}
Soft solids obtained from suspensions of attractive 
colloidal particles are of central relevance in fields 
ranging from food processing to the design of new 
smart materials \cite{new}. In most cases of practical interest 
these solids are not crystalline but gels or soft glasses \cite{suzanne}: they 
arise from the structural arrest of the suspension into a disordered state, 
in which the particles can support external 
stress and therefore behave macroscopically like a solid. 
The mechanical and rheological properties of this soft matter  
depend crucially on its structural features which, in turn, are the result 
of a delicate interplay between the underlying 
thermodynamics and the arrest conditions. Hence, the understanding 
of the mechanisms leading to structural arrest is decisive for many 
technological applications in which colloidal suspensions 
are involved.
Whereas it is clear that in dense suspensions the crowding of the particles 
is the main mechanism leading to structural arrest and yielding, 
suspensions of attractive colloids allow to form soft solids also at low volume fractions, 
where aggregation produces large scale structures 
\cite{lau05,trappe_nature,luca_fd03,segre_01,exp1,edinburgh,bartlett05,
lu06,exp2}. 
At intermediate volume fractions 
(say above $10$-$15\%$ depending on the specific system) possible 
scenarios for the arrest are the crowding of the aggregates acting as the 
new units of a dense glassy system \cite{cates,delgado_na_04,alessio_letter,
reichman_08}, or the geometrically frustrated 
arrangements of locally stable large scale structures 
(such as lamellae or mesoscopic elongated aggregates) 
\cite{decandia_pre06,tarzia_prl06,charbonneau_prl,last}. 
At sufficiently low volume fractions, instead, the structure
of the arrested states, although created via reversible aggregation,
resembles more the open fractal networks typically created by
diffusion-limited aggregation processes \cite{dlca}. 
In this situation, where the arrest seems to be intimately related to 
the network structure, the understanding of the arrest phenomenon is far 
from being reached. 

Our work addresses this last situation, which has received only a 
limited attention in past theoretical and numerical studies.  
Although experiments show that reversible aggregation processes lead 
to the formation of open, persistent network structures,
most of the established models for the attractive effective interactions, 
very successful for many aspects \cite{likos}, fail to produce at these 
low volume fractions network structures that are stable against 
macro- or micro-phase separation. Confocal microscopy images obtained in 
recent experiments \cite{dinsmore_prl06,solomon,royall} demonstrate 
that the gel networks formed at low volume fractions have a distribution 
of the particle coordination number $n$ that is strongly peaked around 
$n \simeq 2,3$, suggesting that the effective interactions include 
also many-body terms. 

As a matter of fact, in these systems there are several possible sources for 
such anisotropies, since the particle surface may not be smooth or the building 
blocks of the gel are not the primary particles but larger aggregates of 
irregular shape \cite{exp2}. Because, up to now, there is no possible 
quantitative, microscopic derivation of effective interaction which accounts 
for these complications, we have recently proposed a phenomenological model containing the relevant 
qualitative ingredients. Following up the idea that directional terms in the 
effective interactions can lead to the formation of an open network that is 
thermodynamically stable, our phenomenological model for the effective 
interactions contains a directional term which introduces a local rigidity 
\cite{delgado_kob_05,delgado_kob_prl,delgado_kob_jnnfm}.
Our aim is not to reproduce one specific experimental system but 
to understand how this type of effective interaction might be relevant for 
the gelation phenomena as observed in diluted suspensions of attractive 
colloidal particles. 

In our model, without imposing a maximal local connectivity 
\cite{zacca}-\cite{martin},
the network structure arises from the balance between the internal energy 
and the entropy, depending on temperature and volume fraction.
In a first study we have shown \cite{delgado_kob_prl,delgado_kob_jnnfm} 
that the formation of the persistent network produces the coexistence, 
in the gel, of very different relaxation processes at different length 
scales: the relaxation at high wave vectors is due to the fast cooperative 
motion of pieces of the gel structure (e.g. the chains connecting two nodes),
whereas at low wave vectors the overall rearrangements of the heterogeneous
gel make the system relax via a stretched exponential decay of the time
correlators. The coexistence of such diverse relaxation mechanisms
is characterized by a typical crossover length which is of the order
of the network mesh size \cite{footnote0}. 
 
In this paper we present a systematic molecular dynamics study of the model 
for different temperatures and volume fractions. 
We will argue that the structural and dynamical features of this model 
should be quite general and could be relevant to different experimental 
gel-forming systems.

In section \ref{model} we give the details of the model and of the
numerical simulations. The structure of the system is analyzed 
in section \ref{structure} and in section \ref{dynamics} we study the dynamics. 
All the results are gathered into a coherent picture
in section \ref{conclu}.
\section{Model and numerical simulations}
\label{model}
\begin{figure}
\begin{center}
\includegraphics[width=0.5\linewidth]{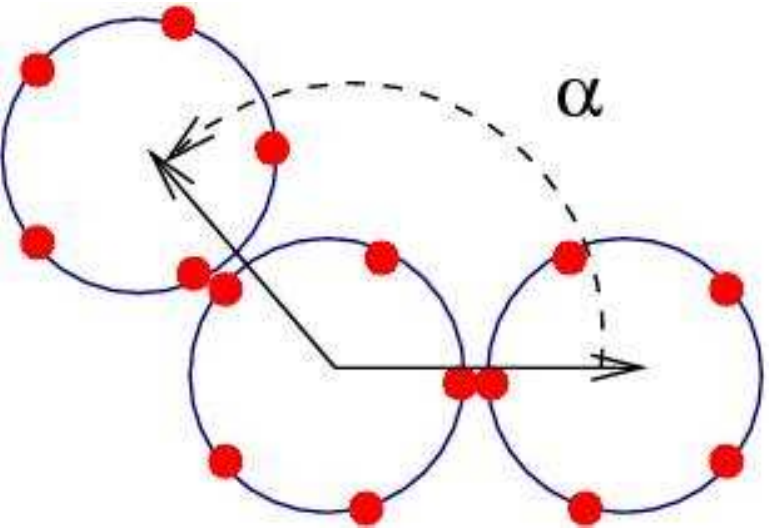}
\caption{The cartoon illustrates the mechanism of the directional interactions:
1) Two particles close within the attraction range and touching with
the {\it sticky} points correspond to an internal energy per particle of $-1$
(in units of $k_B T$). Any other configuration corresponds to an internal
energy per particle of $-0.2$.
2) The angle $\alpha$ among $3$ neighboring particle
is unlikely to be smaller than $70$ degrees.}
\label{fig1}
\end{center}
\end{figure}

We consider identical particles of radius $\sigma$, 
interacting via a potential $V_{eff}$. As mentioned above we seek 
a phenomenological model for the effective interactions which can lead to 
gelation in attractive
colloidal suspensions at low volume fractions. 
For this, for the inter-particle attraction we consider a Lennard-Jones-like
potential of the form 
$V_{LJ}(r_{ij})= 23\epsilon( (\sigma_{LJ}/r_{ij})^{18} - (\sigma_{LJ}/r_{ij})^{16})$,
producing a narrow attractive well ($r_{ij}$ is the distance between the centers of
particles $i$ and $j$).  
To introduce directionality, we have considered an additional 
soft-sphere repulsion modulated by a geometric term, which reduces 
the attractive well of $V_{LJ}(r_{ij})$ unless particles $i$ and $j$ 
are in specific relative configurations. To define such configurations the 
particles are decorated with {\it sticky} 
points occupying the vertices of the icosahedron 
inscribed in the sphere of diameter $\sigma$ 
and centered in the particle position (see Fig.\ref{fig1}).
Between particles $i$ and $j$ the directional effect is then obtained
in the following way:
\begin{eqnarray}
V_{d}(r_{ij}) &=& \frac{\epsilon}{2}\left(\frac{\sigma}{r_{ij}} \right)^{12} \Bigg\{
\left[ \sum_p \left( 1 -
\frac{1}{1+ f({\bf{r}}_{i}-{\bf{r}}_{jp})}
\right)  - 11 \right]
\nonumber \\
& &  + \left[\sum_{p'} \left(1 - \frac{1}{1+
f({\bf{r}}_{j}-{\bf{r}}_{ip'})}
\right)- 11 \right]   \Bigg\}.
\label{ico}
\end{eqnarray}
Here ${\bf{r}}_{i}$ is the position of $i$, ${\bf{r}}_{ip}$
gives the position of the point $p$ on particle $i$.
The sum runs over the $12$ points of respectively $i$ and $j$.
$f({\bf{r}}_{i}-{\bf{r}}_{jp}) = (\frac{(|{\bf{r}}_{i}-{\bf{r}}_{jp}|)^{2}}{d^2})^{8}$
so that, if there is one of these points whose distance from the center of
particle $i$ is small compared to $d$, $V_{d}(r_{ij}) = 0$ and the
additional soft sphere repulsion does not contribute. If this distance is 
larger,
then $V_{d}(r_{ij}) = \frac{\epsilon}{2} (\frac{\sigma}{r_{ij}})^{12}$,
reducing the depth of the original attractive well to $20\%$.
As a consequence, two bonded particles will be subjected to the full attraction
$V_{LJ}$ only if they touch in those specific configurations and 
a reduced attraction otherwise.
The choice of $12$ points allows us to define such relative configurations
 without imposing a local maximum connectivity to the particles: for example, 
the choice of $3$ or $4$ points would anyway lead to the formation of an open 
network interfering with the effect of directionality which is more appropriate 
for the physics of these systems.

The term $V_{\rm d}$ alone is not able to effectively limit 
the functionality of the particles at the volume fraction and temperatures 
considered here \cite{delgado_kob_unp} and will lead to a gel network 
whose structure and dynamics are strongly affected by phase separation kinetics. 
Therefore, in our effective interactions we consider a third term which imposes 
a certain angular rigidity to the bonds $ij$ and $ik$:
\begin{equation}
V_{3}(\hat{r}_{ijk}) = 13.5 \epsilon
\hspace{-0.1cm}\left(\hspace{-0.1cm}\frac{\sigma^{2}}{r_{ij} r_{ik}}\hspace{-0.1cm}\right)^{9} \hspace{-0.1cm}
e^{-\left[\hspace{-0.05cm}\left(\hspace{-0.05cm}\frac{({\bf{r}}_{k}-{\bf{r}}_{i})
\cdot({\bf{r}}_{j}-{\bf{r}}_{i})}
{|{\bf{r}}_{k}-{\bf{r}}_{i}| |{\bf{r}}_{j}-{\bf{r}}_{i}| } - \cos \alpha \hspace{-0.05cm} \right)^{2} / b^{2}
\hspace{-0.05cm}\right]^{2} }
\label{v2}
\end{equation}
making that the angle among the three neighbor particles is unlikely to be
smaller than a certain value $\alpha$ (see Fig.\ref{fig1}).

The resulting interaction potential is
\begin{equation}
V_{eff}= V_{LJ} + V_{d} + V_{3}.
\label{eq1}
\end{equation}
The choice of including both the terms $V_{d}$ and $V_{3}$ has been
made in the spirit of investigating more deeply their relative
contribution to the formation of the open network (and to its dynamics), at a
stage where the presence itself of any directional effect was questioned in the
scientific community and therefore not considered in most of the cases.
Our investigation of the different role of the two terms, albeit partial,
indicates that a simplified version of the model with only $V_{LJ} + V_3$
should lead to very similar structural and dynamical features \cite{kob_sastry_09}.

We have implemented $V_{eff}$ in a constrained molecular dynamics code to
perform micro-canonical simulations, using a suitable combination of
the algorithms RATTLE and SHAKE~\cite{md}. 

Although the model may seem quite complicate, in the following sections it 
will be shown how it undoubtedly picks relevant aspects of the physics it was 
meant to describe and it does contribute to its understanding.

In Eqs.\ref{ico}-\ref{eq1}, the parameters $\sigma_{LJ}$, 
$\sigma$, $b$, $d$ and $\alpha$ can be used to tune the effective 
interactions thus giving rise to a large variety of interesting possibilities 
(e.g. varying the connectivity from chains to compact structures, changing 
the local rigidity of the network...). 
However the desired features of the persistent structure 
do not correspond to a very specific combination of the parameters, 
but to reasonable ranges of values which allow for extensive investigations.
In particular, we have selected the ranges which allow to give rise to the
desired features, like a relatively narrow attractive well,
a connectivity greater than $2$ and a certain angular rigidity.
Here we consider one reasonable choice which guarantees the formation of a 
persistent open network, i.e.
$\sigma_{LJ}=0.922$, $\sigma=1.0$, $d=0.43$, $b=0.34$ and $\alpha=0.4 \pi$.
We have already shown that with these parameters at a volume fraction 
$\phi=0.05$, the static structure of this system at low temperature is 
an open network structure
\cite{delgado_kob_05}, in qualitative agreement with the one of colloidal gels.
Furthermore the system does not show any sign of phase separation in the 
temperature range investigated and its relaxation time increases rapidly with
decreasing $T$, i.e. the static and dynamic properties of the system
are indeed very similar to the ones found in real colloidal gels
~\cite{segre_01}.

In the following sections we will discuss the results of molecular dynamics 
simulations in the micro-canonical ensemble, using a time step of $0.002$. The
unit of time is $\sqrt{m \sigma^2/\epsilon}$, with $m$ the mass of a 
particle.
All the data refers to simulations performed with 8000 particles in cubic
boxes of size $L=37.64, 43.09, 55.10$ in unit of $\sigma$, corresponding 
respectively to particle densities of $\rho = 0.15$, $0.1$, and $0.05$ 
from which we have estimated approximately a volume fraction 
$\phi \simeq 0.075$, $0.05$, and $0.025$.
For every value of $L$ we have studied the system at the temperatures
1.0, 0.7, 0.5, 0.3, 0.2, 0.15, 0.1, 0.09, 0.08, 0.06, 0.055, and 0.05.
In order to do this we have performed the following equilibration protocol:
starting from initial high temperature random configurations,
the system is equilibrated at each temperature by replacing all the
velocities of the center of mass of the particles and the angular velocities
with values extracted from a Maxwell-Boltzmann distribution every
$\Delta$ time steps ($\Delta$ is suitably varied with temperature from 
$10$ to $10^3$ MD steps). 
We have checked that after equilibration the energy is constant, 
showing no significant
drift over the simulation time window. We have monitored independently 
translational and rotational contributions to kinetic energy. Finally, we have 
verified that different one- and two- time autocorrelation functions reach 
the equilibrium behavior, i.e. do not age. 

From these equilibrated
configurations we start the data production. The equilibration time grows 
accordingly to the relaxation time in the 
system. At the lowest temperatures the equilibration procedure required 
up to $2\cdot10^7$ MD steps. 
For each temperature and volume fraction we generated five independent runs, 
over which the results presented here have been averaged.
\section{Structure}
\label{structure}
We investigate the structural changes in the system at the different 
temperatures by means of a comparative study of the density fluctuations, 
of the local connectivity and of the aggregation process.  
\subsection{Static structure factor}
\label{structurea}
After equilibrating the system at different temperatures, we calculate the
static structure factor:
\begin{equation}
S(q) = \frac{1}{N} \sum_{jk} \left( e^{i \bf{q}\cdot(\bf{r_j}- \bf{r_k})} 
\right)
\label{sq}
\end{equation} 
where the values of the modulus of the wave vector $q$ considered are the 
ones compatible with the periodic boundary conditions of the simulations box.
\begin{figure}
\begin{center}
\begin{minipage}{1.\linewidth}
\includegraphics[width=1.0\linewidth]{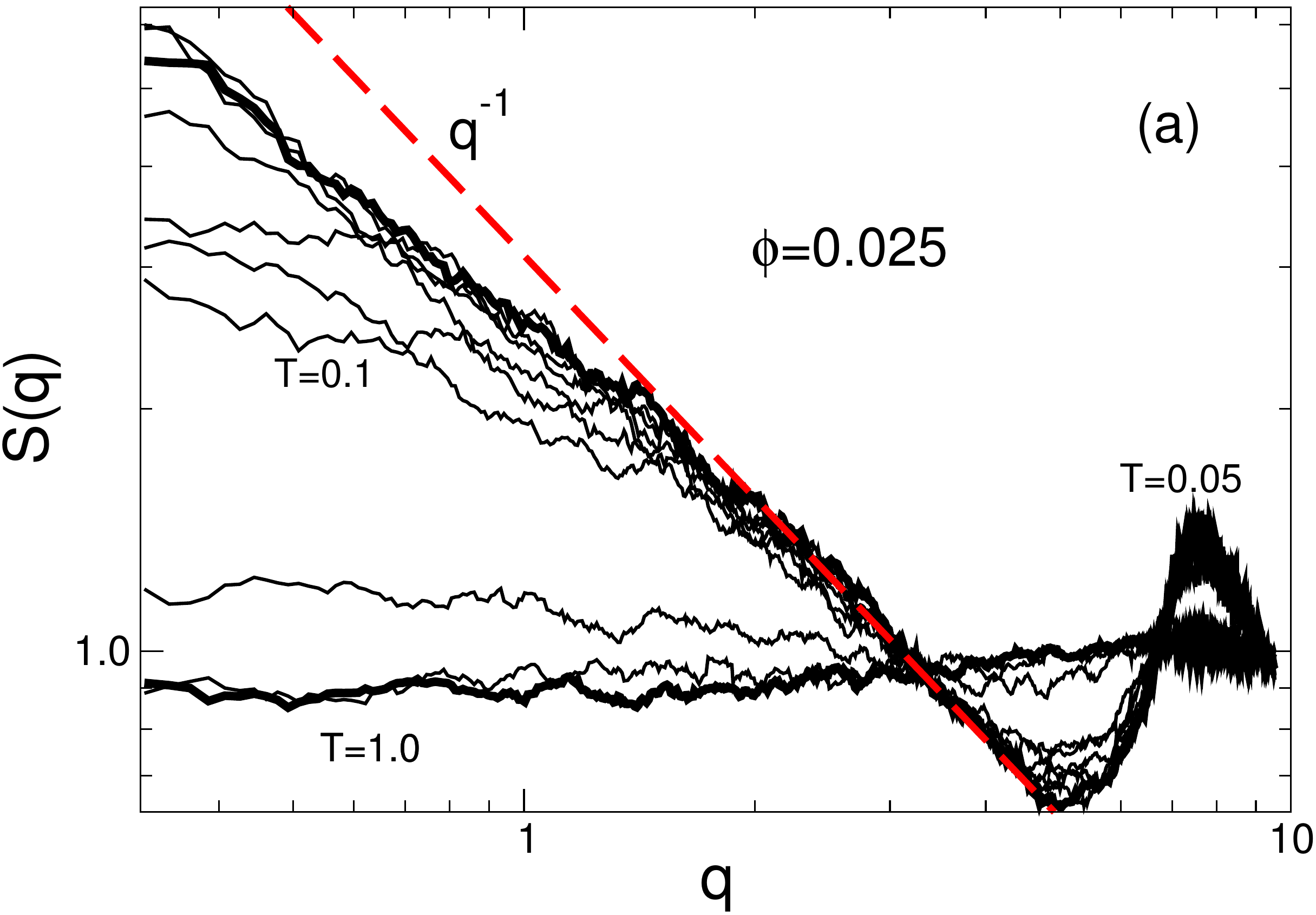}
\end{minipage}
\vspace{0.8cm}
\end{center}
\begin{center}
\begin{minipage}{1.\linewidth}
\includegraphics[width=1.0\linewidth]{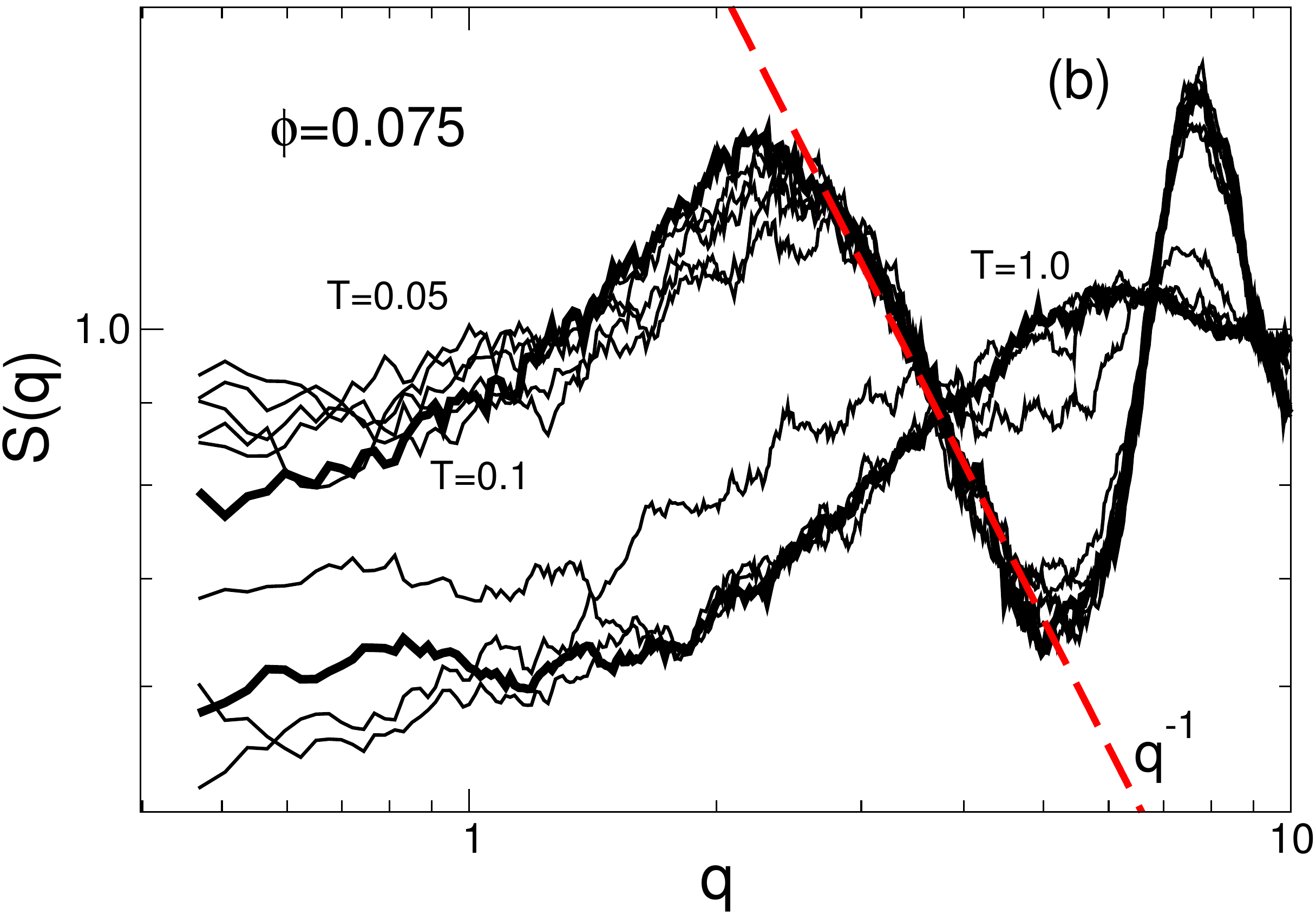}
\end{minipage}
\end{center}
\label{str1}
\caption{The static structure factor as a function of the wave
vector $q$ at volume fraction $\phi=0.025$ (a) and $\phi=0.075$ (b).
In each plot, from bottom to top (at low q), $T= 1.0, 0.5, 0.2, 0.1,
0.09, 0.08, 0.07, 0.06, 0.055, 0.05$. The dashed line indicates the dependence $1/q$.}
\end{figure}

\begin{figure}
\begin{center}
\includegraphics[width=1.0\linewidth]{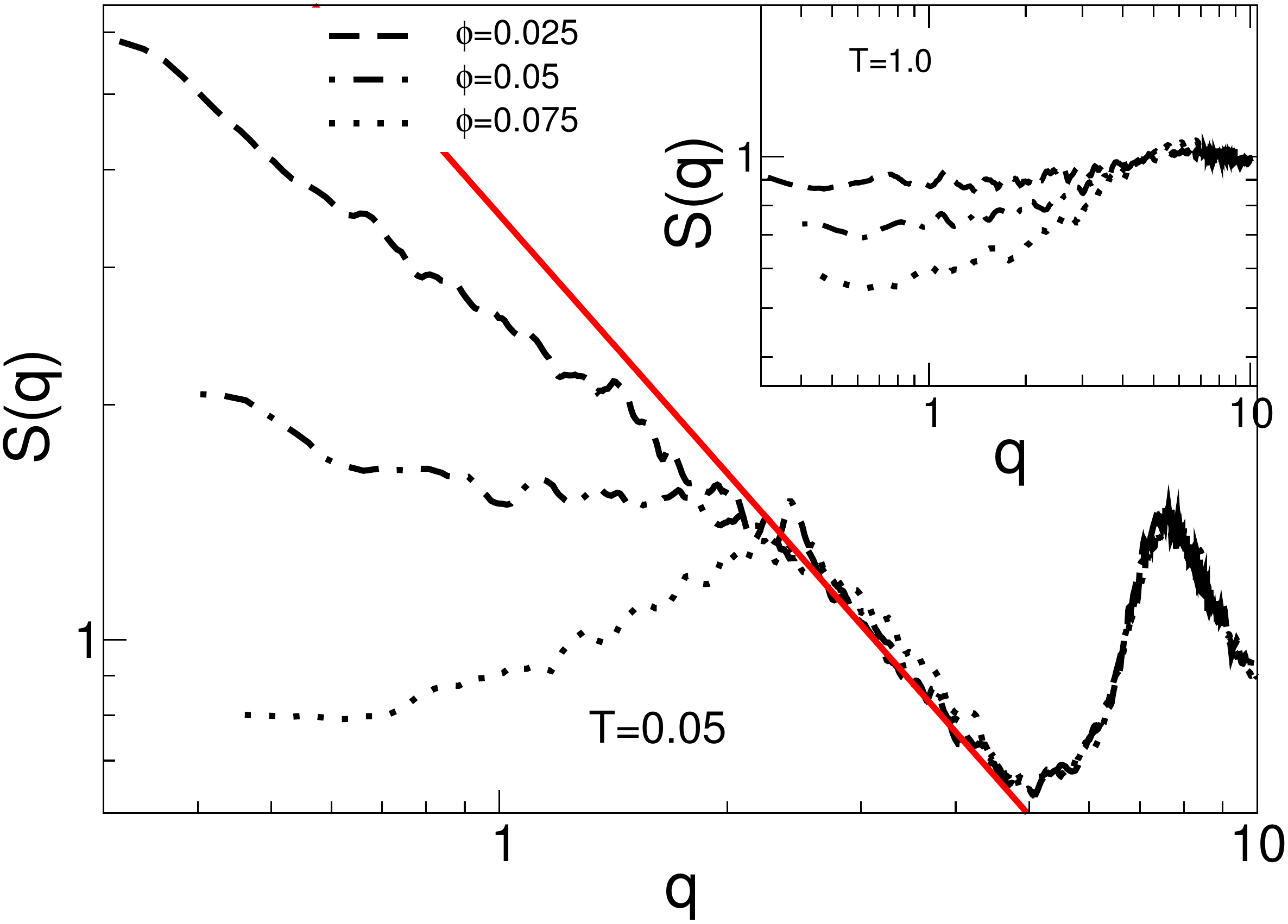}
\caption{
$S(q)$ as a function of the wave vector at
volume fractions $\phi=0.025$, $0.05$, and $0.075$ at 
$T=0.05$ (main frame) and $T=1.0$
(inset). The full line indicate the dependence $1/q$.}
\label{fig4}
\end{center}
\end{figure}

In Fig.~2, we plot 
$S(q)$ as a function of $q$ for
different temperatures, at the volume fractions 
$\phi=0.025$ and $\phi=0.075$. 
The plots show that, upon lowering the 
temperature, major structural changes occur due to aggregation, as indicated 
by the peak located at a wave vector approximately corresponding 
to the first neighbor distance within the potential well i.e. $q \simeq 7.6$, 
and, more 
importantly, by the simultaneous onset of spatial correlations at low 
wave vectors.
At high wave vectors the change of $S(q)$ from high to low temperatures
looks qualitatively the same at different $\phi$.  
The spatial correlations at low wave vectors, and therefore the
mesoscopic and large scale structures, seem instead to depend strongly 
not only on $T$ but also on $\phi$.
We observe that the data of Fig. 2 show no sign of
a phase separation, which would be indicated by a 
rapidly growing peak at low wave vectors. 
Instead we see at low $q$ the onset of spatial correlations 
extending over different length scales, suggesting the formation 
of an extended disordered structure.
We compare in Fig.~\ref{fig4} the $S(q)$ at the different $\phi$:
at high temperature (inset of the figure), 
spatial correlations in the system are mainly temperature-controlled, 
i.e. depend only weakly on $\phi$.
At the lowest temperature (main frame), instead, 
space correlations are dominated by temperature up to length scales
of the order of a few particle diameters. In this region of $q$, 
i.e. $2.0 \leq q \leq 5.0$, $S(q)$ shows the same $ \propto 1/q$ 
dependence for the three values of volume fractions considered. 
This clearly indicates the presence of mesoscopic 
elongated structures (chains) as a result of the aggregation driven 
by the potential energy.
Finally, beyond length scales of the order of $3$-$4$ particle diameters 
(i.e. $q\leq 2.0$), the strength and range of spatial correlations are 
strongly controlled by the volume fraction. 

If we consider now $S(q)$ in analogy with the static structure factor of 
a polymer chain solution~\cite{rubinstein}, 
we can interpret the mesoscopic length scales 
$2.0\leq q \leq 7.0$ as an {\it intra-molecular} 
regime for spatial correlations due to the elongated aggregates and the 
macroscopic scales $q\leq 2.0$ (i.e. for lengths up to the simulations box 
linear size) as an {\it inter-molecular} regime. Following this description, 
the {\it intra-molecular} regime is strongly controlled by the interaction
potential (and therefore by temperature), and extends up to a length-scale
which gives a rough estimate of the persistence length of the elongated 
aggregates ($\geq$ 3-4 particle diameters).
Beyond the persistence length, the $S(q)$ captures the {\it inter-molecular}
regime: the fact that this regime depends quite dramatically on the volume 
fraction even in the small interval here considered ($\Delta\phi = 0.05$) 
suggests that the linear size of the aggregates extends well beyond their 
persistence length. 

To conclude this section, we make two 
remarks. First, we notice that at the lowest volume fraction $\phi=0.025$,
one recognizes a power law regime in 
$S(q)$ at intermediate and low $q$. It is clear that 
this cannot be interpreted as a fractal regime, since the exponent would 
be less than 1.0. It should instead correspond to the crossover 
from the {\it inter-molecular} regime to an eventual fractal regime which is only 
detectable for a larger system. This will be confirmed by the
analysis of the structure in the following sections. 
Second, we observe that, at the highest volume fraction, $\phi=0.075$,
$S(q)$ displays at low $T$ a well defined pre-peak at a wave vector 
$q \simeq 2.0$. In network-forming ionic liquids, this kind of pattern 
in scattering intensity is typically associated to the presence of a
network structure \cite{ionic}. In gelling colloidal suspensions 
it has also been related to the presence of stable clusters of a typical size of
$3$-$6$ particles
~\cite{segre_01,stradner} and its connection to gelation is debated.
In fact, $S(q)$ can give information only on the typical distances 
between pairs of particles occurring in the structure and does not 
allow on its own to discriminate between pairs of particles in the same 
aggregate or from different aggregates. Here, the pre-peak 
arises from the crossover between the {\it intra-molecular} and the 
{\it inter-molecular} regime at this volume fraction due to two competing 
effects: pairs of particles in the same large 
aggregate (chain) will be typically separated at least by a distance of 
$3$-$4$ particle diameters, due to the local rigidity;
pairs of particles separated by larger distances, belonging or not to 
the same aggregate, will be limited instead upon increasing the volume fraction. 
With numerical simulations we have the possibility to access 
further structural features, to complement the information 
obtained through $S(q)$. Therefore in the following we exploit it by 
analyzing the local connectivity of particles, as well as the cluster 
size distribution, as a function of the temperature and of the volume fraction.
\subsection{Connectivity}
\label{structureb}
\begin{figure}[h]
\begin{center}
\includegraphics[width=1.0\linewidth]{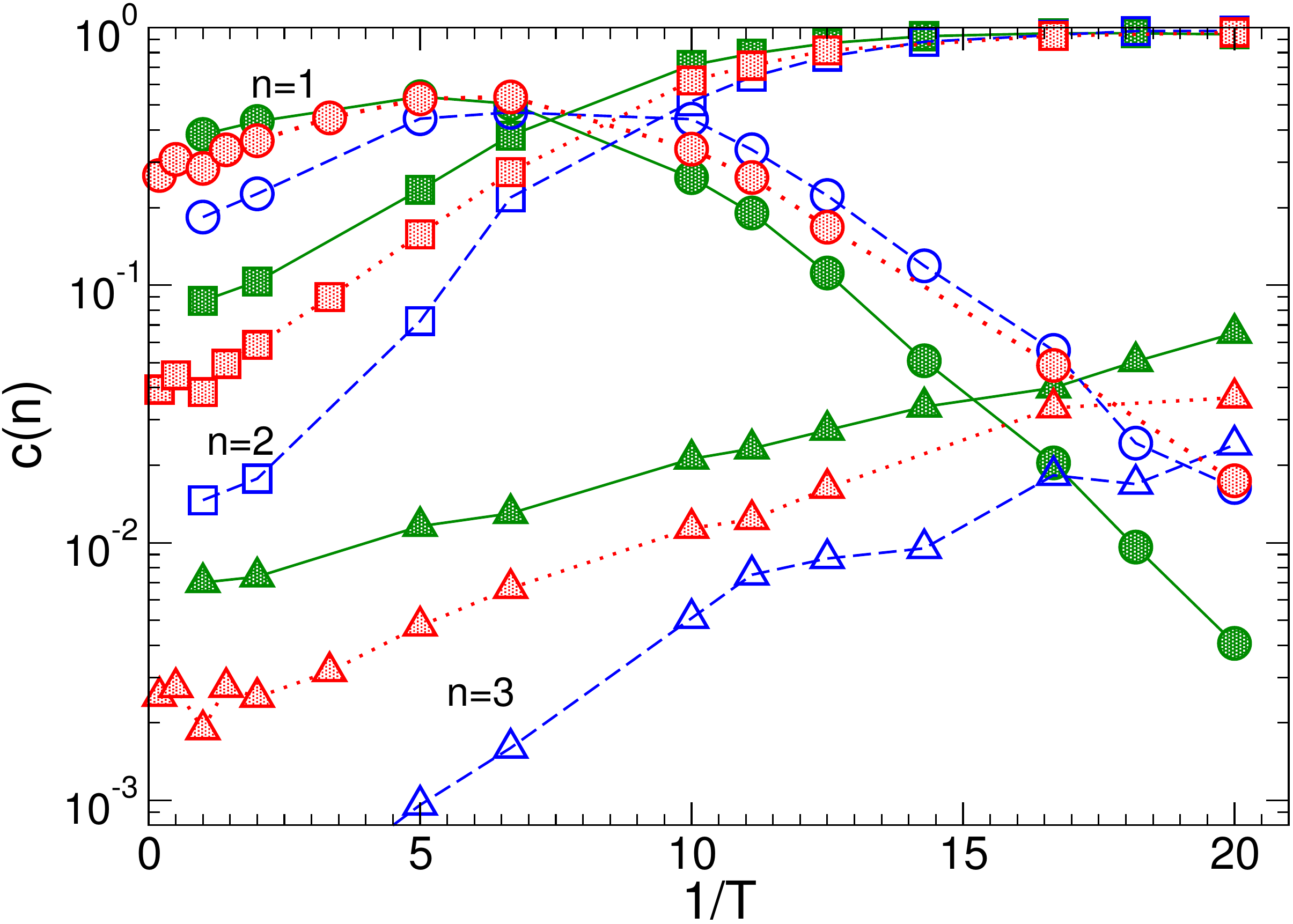}
\caption{
Coordination number $c(n)$ as a function of inverse temperature.
The different symbols correspond to $n=1$ ($\circ$),
$n=2$ ($\square$) and $n=3$ ($\triangle$). The different connecting lines
correspond to $\phi=0.025$ (dashed), $\phi=0.05$ (dotted) and $\phi=0.075$
(full).
}
\label{coord}
\end{center}
\end{figure}

We define the coordination number $c(n)$ as the fraction of particles 
that have exactly $n$ neighbors. Two particles are considered to be 
neighbors if their distance is less that $r_{\rm min}$=1.1, the
location of the first minimum in the radial distribution function.
By means of $c(n)$ we characterize the change in the topology of the 
structure occurring if $T$ is decreased.
In Fig.~\ref{coord}, $c(n)$ is plotted as a function of inverse 
temperature for different values of $n$ and for the different volume 
fractions considered here.

As a general result, at high temperatures the vast majority of the 
particles are isolated, ($n=0$ not shown), and the fraction of dimers, $n=1$,
is relatively high ($20-30\%$). With decreasing $T$ this fraction 
initially increases, but for $T\leq 0.2$ it starts to rapidly decrease 
with decreasing $T$. Our data shows that this happens because particles
start to form larger structure which are mainly chains:
the fraction of particles with exactly two nearest neighbors increases 
rapidly and these (local) configurations become by far the most prevalent
ones at low $T$. Last but not least also the fraction of particles with $n=3$
neighbors increases with decreasing $T$. 
Therefore the curves
of Fig.~\ref{coord} indicate that upon decreasing $T$ the particles form 
chains which are connected by bridging points or nodes ($n=3$) to form 
an open network. It is interesting to remark that 
at the lowest temperature there are practically no free particles.
The fraction of dimers ($n=1$) is minor and it significantly decreases with 
increasing the volume fraction. The fraction of particles with $n=2$ 
does not strongly depend on $\phi$ in the range of values explored, whereas 
the fraction of particles with $n=3$ monotonically increases with $\phi$: this 
suggests that the formation of chains is strongly driven from the interaction, 
whereas the increasing steric hindrances between large aggregates 
due to increasing $\phi$ will favor the 
formation of junctions between the chains by counterbalancing the angular 
repulsion $V_{3}$.

\subsection{Clusters}
\label{structurec}
We have monitored the aggregation process and characterized the structure 
of the system on large length scales using $n(s)$, the number of clusters 
that have exactly $s$ particles. We define that a particle belongs to a
cluster if its distance from at least one member of the cluster is less
than $r_{\rm min}$, the location of the first minimum in the radial 
distribution function. 
\begin{figure}
\begin{center}
\begin{minipage}{1.\linewidth}
\includegraphics[width=1.0\linewidth]{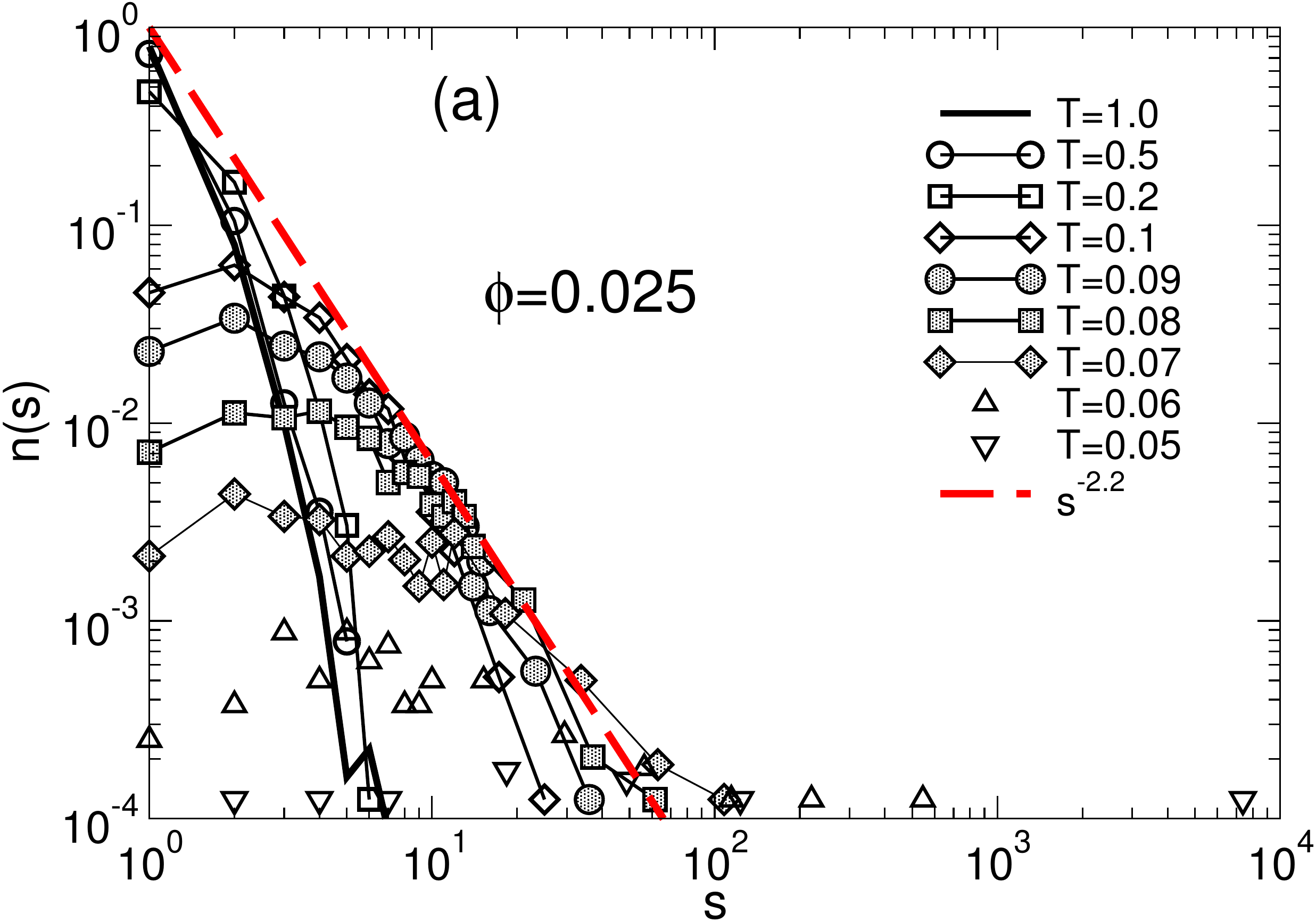}
\end{minipage}
\vspace{0.8cm}
\end{center}
\begin{center}
\begin{minipage}{1.\linewidth}
\includegraphics[width=1.0\linewidth]{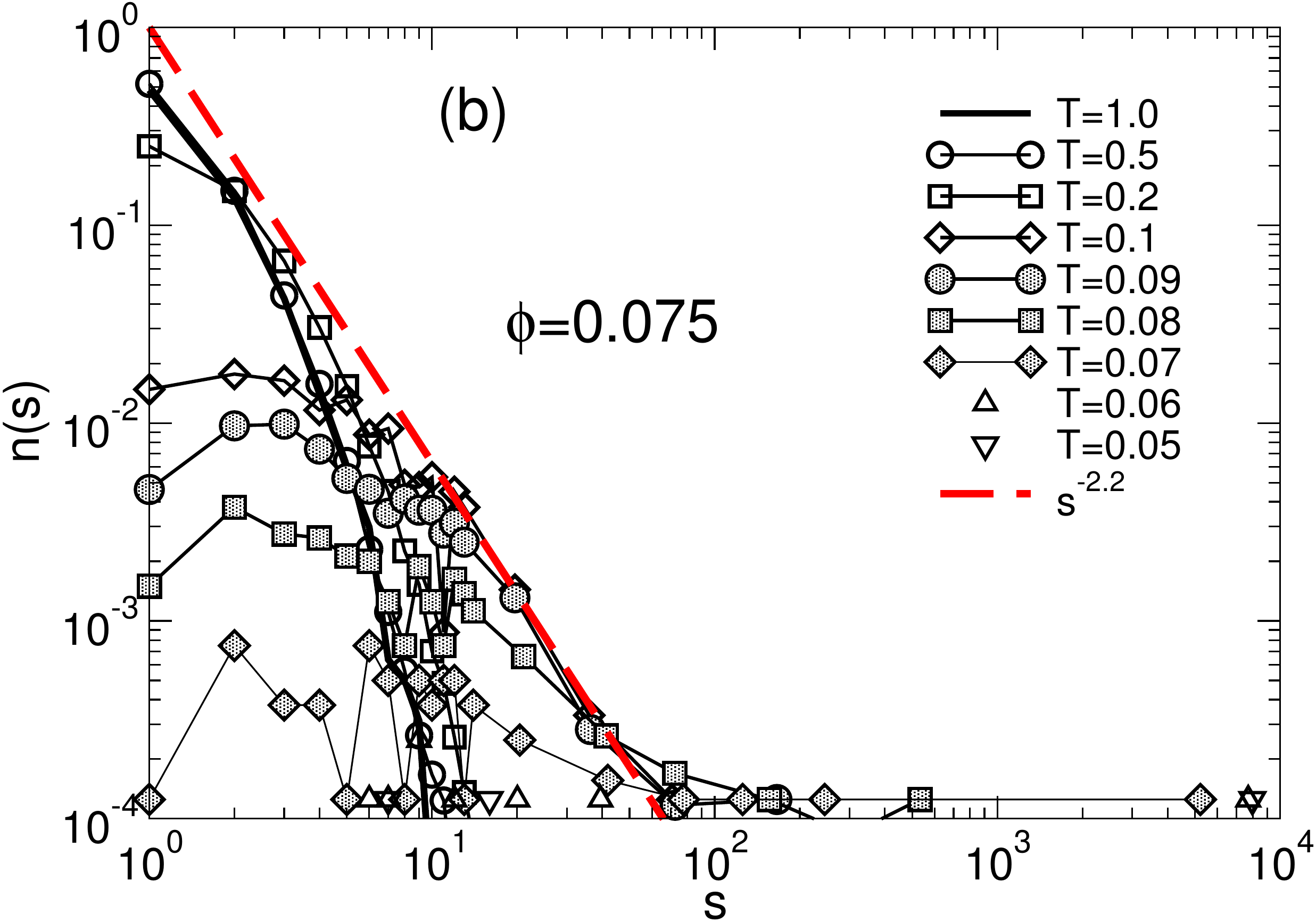}
\caption{Cluster size distribution $n(s)$ for the different temperatures at
volume fraction $\phi=0.025$ (a), and $\phi=0.075$ (b). The dashed line 
corresponds to $s^{-2.2}$. 
}
\end{minipage}
\label{fig6}
\end{center}
\end{figure}
This distribution is shown in Fig.~6 
for all temperatures and volume fractions investigated.
For the different volume fractions, at high temperatures ($T\geq 0.3$) 
the distribution nicely follows an exponential law, a behavior 
that corresponds to the random formation of transient clusters of non-bonded 
particles at low densities. Around $T=0.1$ the shape of the distribution 
starts to change strongly, indicating that dimers are the most probable 
clusters. At the same time, $n(s)$ also starts to show a tail at large $s$, with 
clusters sizes of the order of a few hundreds particles. At lower 
temperatures, $n(s)$ displays a power law regime for high values of 
$s$, with a crossover point that moves to larger $s$ with decreasing $T$. 
This regime is compatible with an 
exponent around $-2.2$ (the dashed line in the plots of Fig.~6)
in agreement with random percolation~\cite{stauffer}. 
By comparing the data at different volume fractions we observe that the 
temperature at which the power law tail appears increases with 
increasing volume fraction. The percolation threshold, estimated  
as the temperature where $50\%$ of the configurations percolate, monotonically 
increases from $T\simeq0.08$ at $\phi=0.025$ to $T\simeq0.14$ at $\phi=0.075$. 
This is coherent with the information obtained from
$c(n)$, in Fig.~\ref{coord}: the relative higher amount 
of bridging between chains upon increasing volume fraction corresponds to 
a higher probability for the aggregating structure to percolate at the same 
temperature. Finally, we also conclude from Fig.~5 that, 
once a percolating cluster is formed, the 
particles rapidly aggregate into a unique interconnected structure. This is in 
agreement also with the data of Fig.~\ref{coord} showing that, at the lowest 
temperature, there are practically no free particles and the fraction of 
particles with only one neighbor is negligible, i.e. limited to dangling ends. 
This feature seems qualitatively the same here and in simulations 
performed on systems of smaller size ($1000$ particles), suggesting that
it is not much affected by the system size. 
\subsection{Summary and discussion of the structural analysis.}
\label{structured}
The information contained in Figs.~2-5 allows us to reach 
a convincing interpretation of the scattering patterns of Fig. 2. 
The behavior of $c(n)$ indicates that the mesoscopic length scale discussed above 
corresponds to the formation of semi-flexible chains 
whose persistence length is directly related to the features of the
effective interactions. The {\it inter-molecular} regime, strongly dependent
on the volume fraction, corresponds instead to the formation of a 
network structure due to bridging between different chains. 
Depending on the
the volume fraction, the extended percolating structure will 
have different amount of nodes (i.e. bridging connections) 
and therefore, presumably, significantly different mechanical properties.
Interestingly, this corresponds to very different patterns in $S(q)$ at 
low wave vectors: from the display of a wide crossover for the softest
structure, to the presence of a well defined pre-peak (i.e. a peak at 
wave vectors corresponding to distances significantly larger than the particle 
diameter). Differently from the assumption often made that such peak of 
correlation corresponds to the presence of stable finite clusters,
here this pattern indicates instead the persistence length of our elongated 
aggregates. Our interpretation 
suggests that, in a case like this, the position of the peak could be therefore 
related to the mechanical properties of the network. 
Finally the aggregation process seems to be characterized by the fact that,
once that a percolated structure is formed, particles
aggregate rapidly into a single network structure and small clusters as well 
as free particles completely disappear. This feature, which might be 
distinctive of aggregation induced colloidal gelation, significantly modifies 
the cluster size distribution and has some consequences also on the dynamics.  

The general understanding of the structure formation discussed here will be 
relevant to the analysis of the dynamics, which is 
performed in the following section. 
\section{Dynamics}
\label{dynamics}
We characterize the dynamics in our model by means of different quantities: 
The mean squared 
displacement of the particles, 
\begin{equation}
\langle \Delta r^{2}(t) \rangle = \frac{1}{N} \langle \sum_{j=1}^{N} 
(\mathbf{r}_{j}(t) - \mathbf{r}_{j}(0))^{2} \rangle,
\label{msd}
\end{equation}
and the incoherent scattering function
\begin{equation}
F_s(q,t)=\frac{1}{N}\sum_{j=1}^N \langle \exp[i {\bf q} \cdot
({\bf r}_j(t)-{\bf r}_j(0))] \rangle.
\label{self}
\end{equation}
In order to fully understand the role of the structure formation in the
dynamics we also calculate time correlations of bonds and nodes, defined above,
in the following way:
\begin{equation}
C_b(t) = \frac{\sum_{ij}
\left[ \langle n_{ij}(t)n_{ij}(0)\rangle-\langle n_{ij}\rangle^2\right]}%
{\sum_{ij}\left[\langle n_{ij}^2\rangle-\langle n_{ij}\rangle^2\right]},
\label{bond}
\end{equation}
where $n_{ij}(t)=1$ if particles $i$ and $j$ are linked at time $t$ and
$n_{ij}(t)=0$ otherwise. For the nodes (i.e. particles connected to 
three other particles) $C_{3b}(t)$ is defined as
\begin{equation}
C_{3b}(t) = \frac{\sum_{i}
\left[ \langle n_{3i}(t)n_{3i}(0)\rangle-\langle n_{3i}\rangle^2\right]}%
{\sum_{i}\left[\langle n_{3i}^2\rangle-\langle n_{3i}\rangle^2\right]},
\label{bond3}
\end{equation}
where $n_{3i}(t)=1$ if particle $i$ is a node at
time $t$, $n_{3i}(t)=0$ otherwise.
\subsection{Characteristic time scales from time autocorrelation functions}
\label{dynamicsa}
From the time correlation functions introduced above we calculate the 
characteristic times 
$\tau_{s}(q)= \int F_{s}(q,t) dt$, 
$\tau_{b}= \int C_b(t) dt$ and
$\tau_{3b} = \int C_{3b}(t) dt$, where suitable cut-offs have been used to 
evaluate the time integrals.
\begin{figure}
\begin{center}
\begin{minipage}{1.\linewidth}
\includegraphics[width=1.0\linewidth]{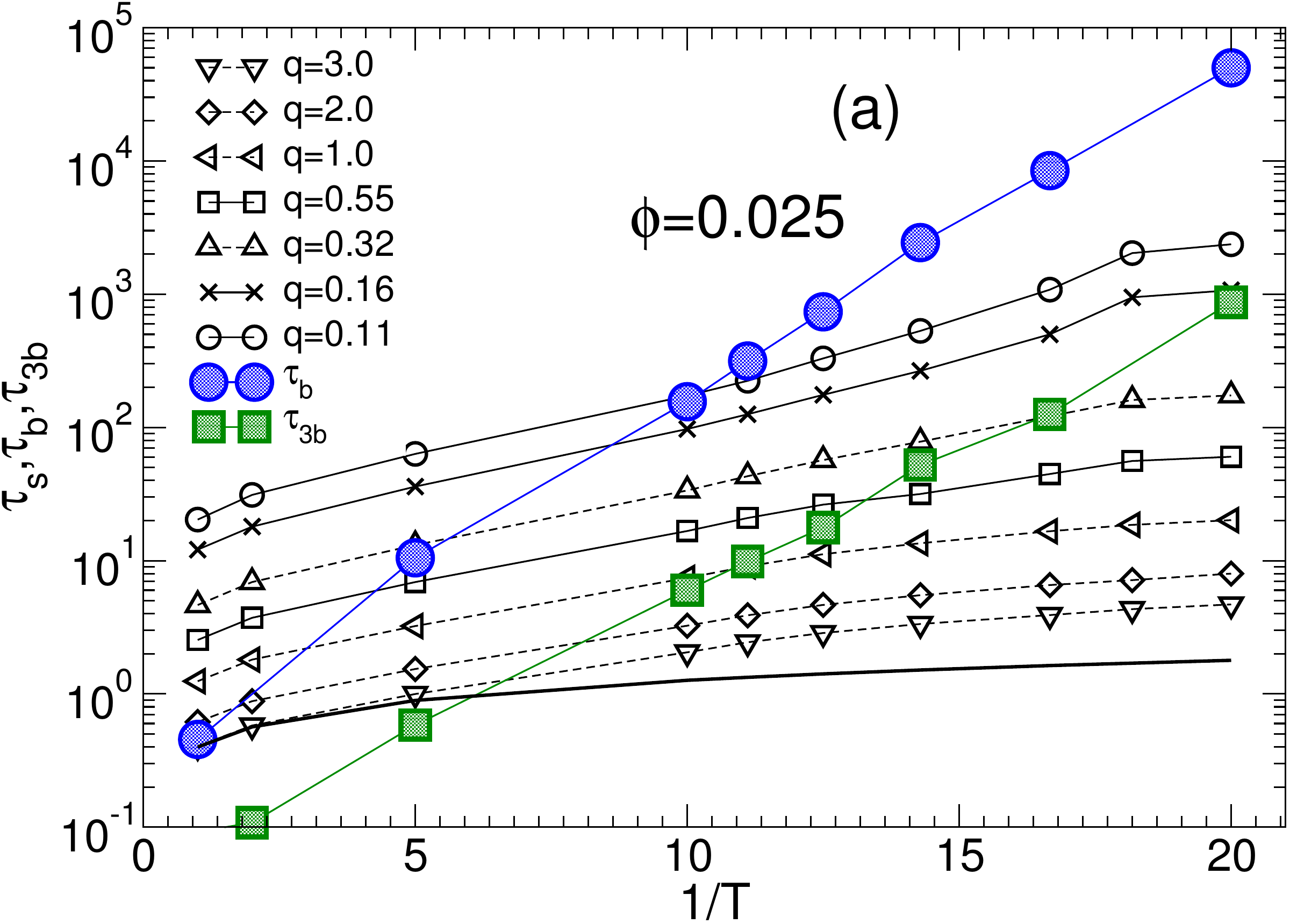}
\end{minipage}
\vspace{0.8cm}
\end{center}
\begin{center}
\begin{minipage}{1.\linewidth}
\includegraphics[width=1.0\linewidth]{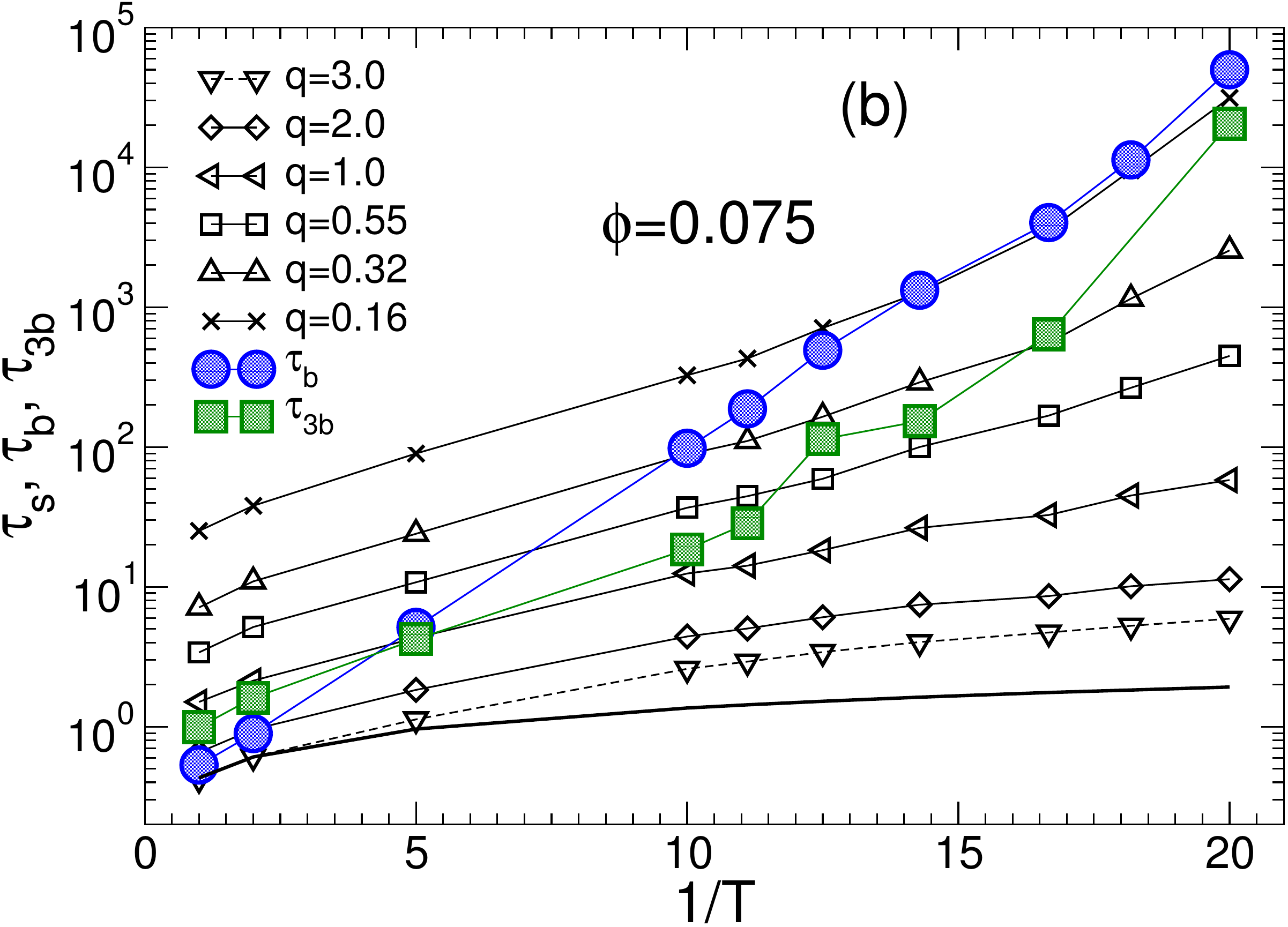}
\end{minipage}
\caption{
Arrhenius plot 
of the relaxation time $\tau_s(q,T)$ as determined from the
self-intermediate scattering function $F_s(q,t)$. The solid line is a
fit to the high $T$ data for $q=3.0$ of the form $\tau={\rm const.}/
\sqrt{T}$. The open symbols correspond to different wave-vectors. The dark 
circles and squares are $\tau_b$ and $\tau_{3b}$, respectively.
}
\label{tscales2}
\end{center}
\end{figure}
Fig.~\ref{tscales2} shows the dependence of these different characteristic 
times on the inverse temperature at volume fraction $\phi=0.025$ (a) and 
$\phi=0.075$ (b). 
At high $T$ the relaxation dynamics at large $q$ can be
approximated by the function $F_s(q,t) = \exp[-Tq^{2} t^{2} /(2m)]$, which
gives $\tau_s= (\sqrt{\pi m /2T})/q$ (solid line).
At these temperatures, bonds are not persistent enough to create long living structures. 
Upon lowering the temperature, bond lifetime becomes the longest 
relaxation time scale. Moreover, at the lowest temperatures
the lifetime of the network nodes becomes 
comparable to the relaxation times at low wave vectors. 
That is, the disordered network structure is persistent enough to affect
the dynamics at large length scales. From the figure we also conclude 
that, whereas
the bond lifetime is fairly insensitive to the volume fraction, the node 
lifetime displays a stronger dependence on $\phi$: the persistence of the nodes
apparently depends on the structural features of the network, which in turn, 
as seen above, depends on $\phi$. This already suggests, as 
discussed more in the following, that the network formation is 
related to the onset of cooperative dynamical processes.
We also observe that at 
$\phi=0.075$, as also found at $\phi=0.05$ \cite{delgado_kob_05}, 
the relaxation time 
$\tau_{s}(q,T)$ at the 
smallest wave vector $q_{min}\simeq 0.16$ shows a temperature dependence which 
becomes stronger 
than Arrhenius at low $T$, whereas this does not happen at the lowest 
volume fraction $\phi=0.025$. 
Upon increasing $\phi$, the increase of node lifetime is apparently 
stronger. Again, this suggests that, whereas the bond breaking 
is an activated process mainly controlled by the potential energy parameters, 
the node lifetime is actually affected by the features of the different
structure at different volume fractions. From the information gained 
from $S(q)$ and our analysis of section \ref{structure}, we  
expect the large lengths scale properties of the structure to play a 
relevant role in the node lifetime. We will come back to these considerations
in the following, when analyzing the decay of time correlations of particle 
displacement, bonds, and nodes.
\subsection{Mean squared displacement}
\label{dynamicsb}
\begin{figure}
\begin{center}
\includegraphics[width=1.0\linewidth]{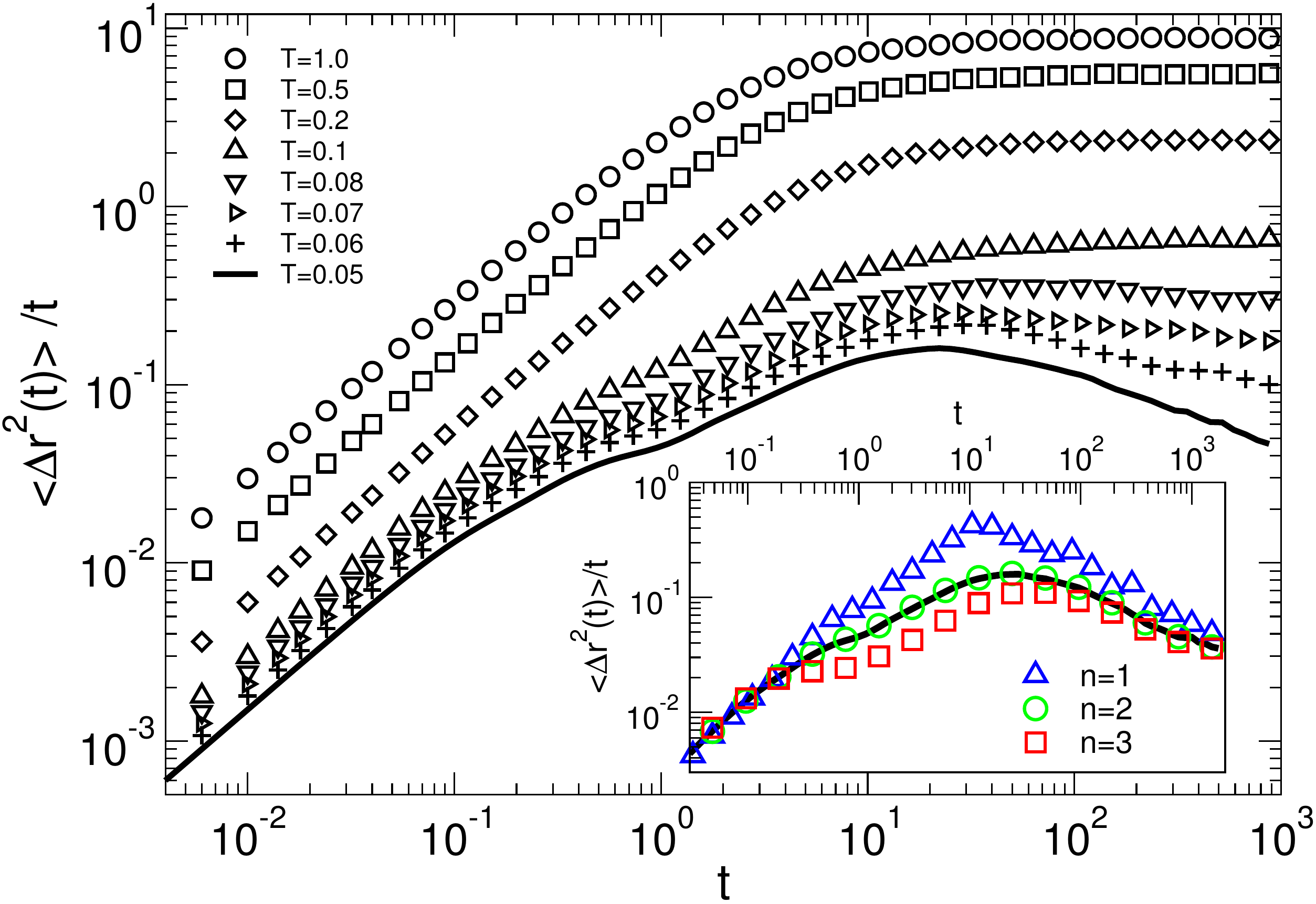}
\caption{
Main frame: $\langle \Delta r^{2}(t) \rangle /t$ as a function of time
at volume fractions $\phi=0.075$ for different temperatures. Inset:
comparison of the data at $T=0.05$ of the main frame with the same 
quantity (full line) calculated for nodes, chains particles and dangling ends.}
\label{msd1}
\end{center}
\end{figure}

\begin{figure}
\begin{center}
\includegraphics[width=1.0\linewidth]{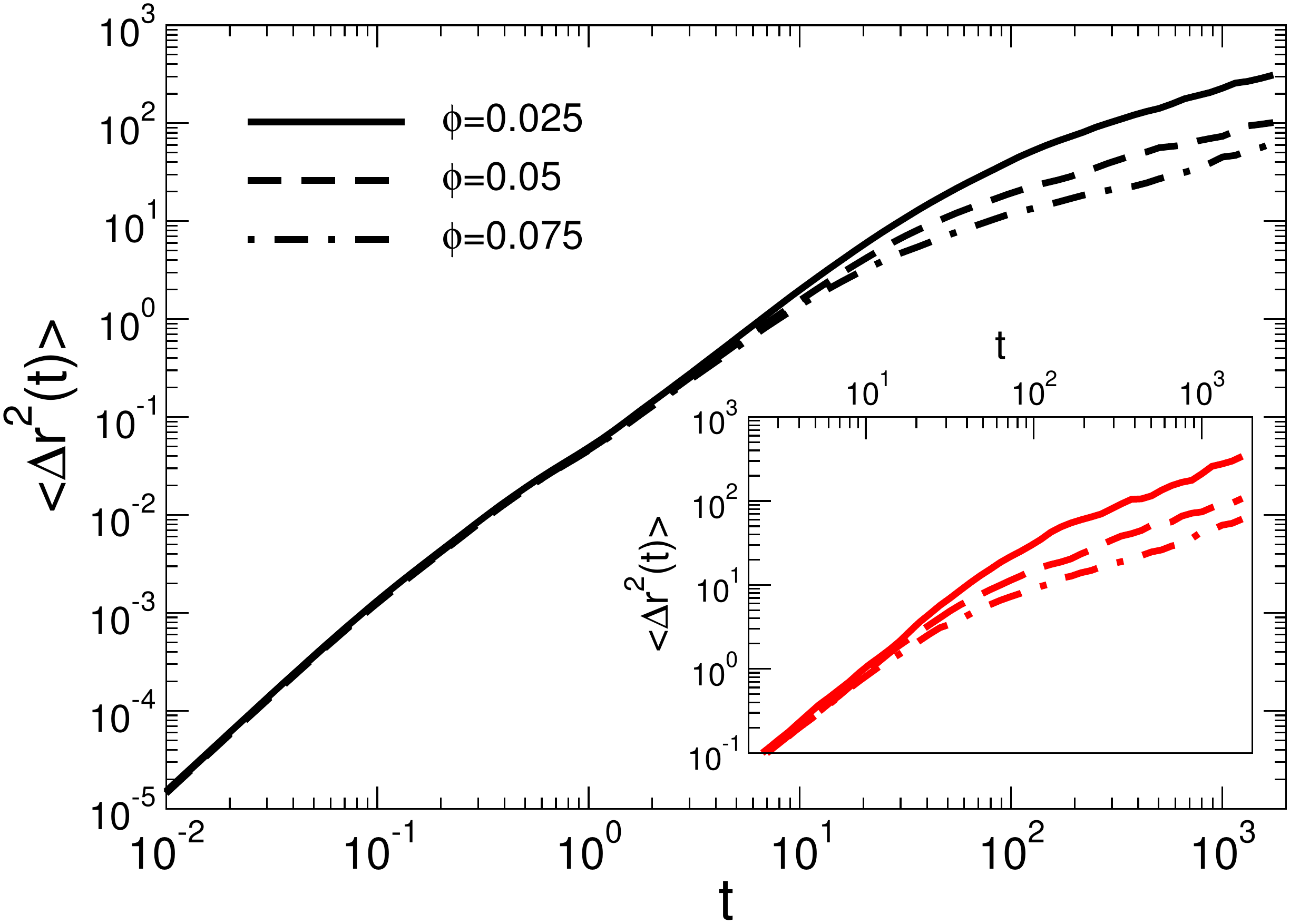}
\caption{
Main frame: Particle MSD as a function of time
at $T=0.05$. Inset: MSD for the nodes at $T=0.05$ for different volume fractions.}
\label{msd2}
\end{center}
\vspace{0.5cm}
\end{figure}

At high temperatures, the time dependence of the particle mean squared 
displacement (MSD) 
crosses over from the ballistic $t^{2}$-dependence at short times to the 
diffusive, i.e. linear $t$-dependence. 
Upon lowering the temperature we 
observe the onset of a more complex behavior at a temperature $T\simeq0.1$, 
at which the attraction starts to create persistent bonding. 
In Fig.~\ref{msd1} we plot 
$\langle \Delta r^2(t)\rangle/t$ at $\phi=0.075$ for the different temperatures 
(main frame). This plot clearly shows that, upon lowering the 
temperature, two different localization processes arise.
The first one, at $t\approx 1$, corresponds to
a localization length $\approx 0.2$, which shows a relatively weak dependence on
further lowering temperature: this is the onset of the caging regime in
which a particle is temporarily trapped by its nearest neighbors,
due to the formation of 
bonds~\cite{delgado_kob_05,delgado_kob_prl,delgado_kob_jnnfm}. 
At the lowest temperatures, the localization process arising at times 
$t\approx 10^3$ corresponds to a localization distance of the order of 
10 particles diameter, i.e. comparable to the size of the mesh
in the disordered network structure. In the inset, the data at the lowest 
temperature $T=0.05$, where practically all the particles belong to a single 
percolated cluster, are directly compared to the contribution coming from 
the nodes of the network (particles with coordination number $n=3$), from
particles belonging to chains ($n=2$) and from the dangling ends ($n=1$). 
This $n$-dependence gives two main indications: the first localization 
process strongly limits the mobility of the network nodes, but the overall MSD 
is dominated by the chain mobility; the second localization process,
present only at the lowest temperature where the lifetime of the network 
nodes becomes comparable to the longest characteristic time scale, strongly
limits the large scale mobility, dominating the MSD. 
The particle MSD for the different $\phi$ at $T=0.05$, i.e. when the network 
is fully developed, is plotted in Fig.\ref{msd2} (main frame) 
as a function of time. Here it is 
clear that, whereas the first localization process ($t \approx 1.$) is very 
weakly dependent on $\phi$, the volume fraction dependence of the second 
localization process ($t\approx 10^{3}$) is much stronger. This feature
is also present if one isolates the contribution coming from the 
nodes of the network (inset of Fig.\ref{msd2}) at the different volume fractions.
These observations are coherent with the volume fraction dependence of 
the structure on different length scales, as shown by the $S(q)$ in 
section \ref{structure}. 

To get a deeper understanding of the correlated particle motion, it is useful 
to monitor the deviation from a Gaussian distribution for the particle 
displacements, which is quantified by the non-Gaussian parameter
\begin{equation}
\alpha_2(t)=\frac{3 \langle \Delta r^4(t)\rangle} {5\langle\Delta r^2(t)
\rangle^2}-1 .
\label{alpha2}
\end{equation}
\begin{figure}
\begin{center}
\includegraphics[width=1.0\linewidth]{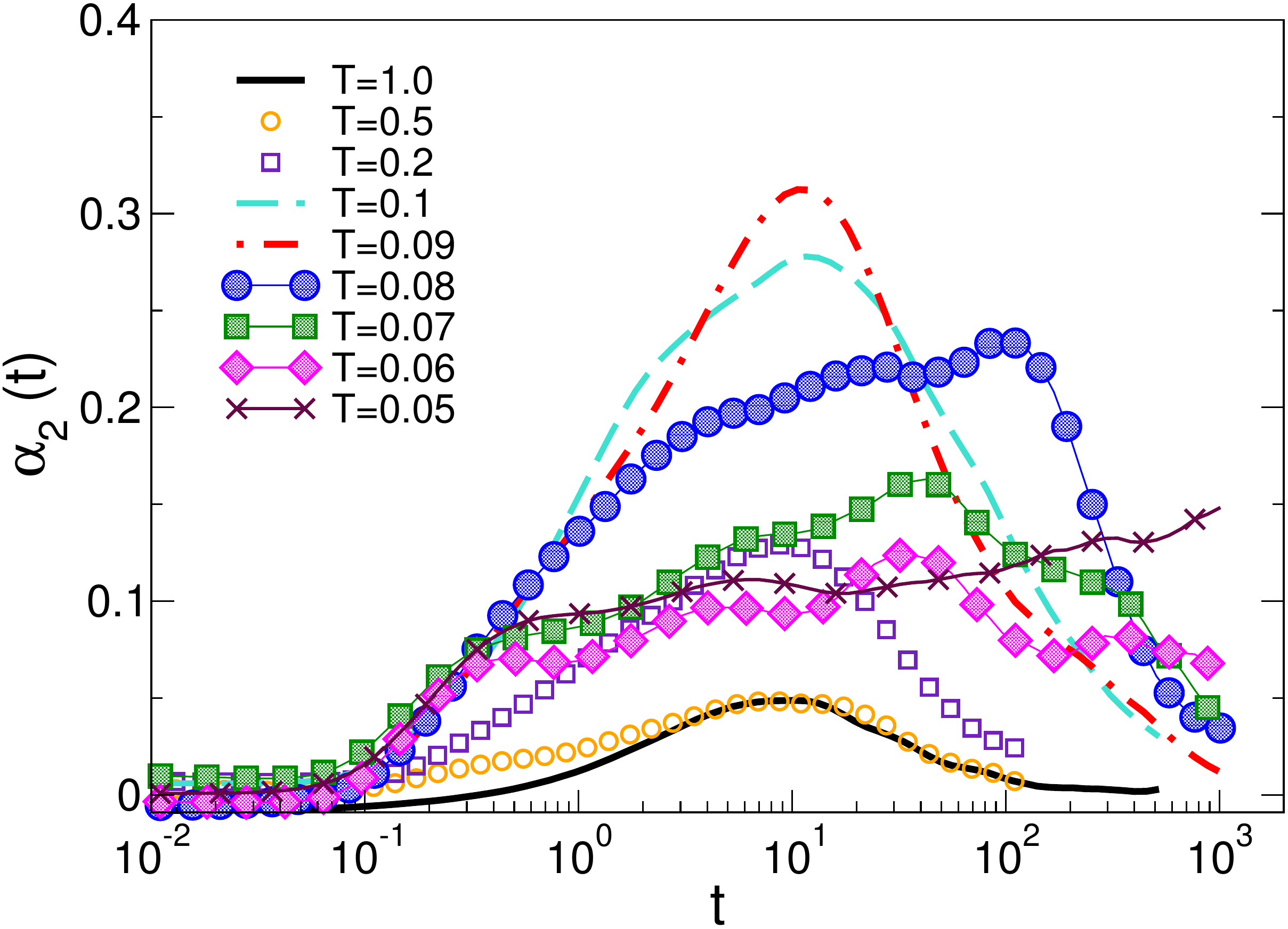}
\caption{Non-Gaussian parameter $\alpha_2(t)$ as a function of time 
for $\phi=0.05$ at different temperatures.
}
\label{alpha2_1}
\end{center}
\end{figure}
In Fig.~\ref{alpha2_1} we show $\alpha_2(t)$ as a function of time 
at different temperatures and volume fraction $\phi=0.05$: at high temperature
there is a small peak in $\alpha_2(t)$ arising at the crossover between the 
ballistic and the diffusive regime. Upon decreasing temperature, this peak 
strongly increases, indicating that the first localization process
induces increasing non-Gaussian contributions to the 
crossover into the diffusive regime. This is actually similar to what is 
typically observed in the non-Gaussian parameter for supercooled 
liquids at the onset of the caging regime. At even lower 
temperature, at which the network starts to form, one can clearly see a 
qualitative change in the non-Gaussian contribution to particle motion: 
the {\it glassy} peak intensity at $t \approx 10$ 
decreases and the maximum of $\alpha_2(t)$ 
apparently moves towards much longer times, occurring 
approximately after the second 
localization process. 
This indicates that, once the aggregation leads towards the network, 
the main non-Gaussian contribution to particle motion is due to the 
second localization process, i.e. on length scales of the order of network mesh 
size. 
This indicates that there is in fact a different glassy 
regime of the relaxation dynamics which is set in by the formation of 
the persistent network and whose onset is therefore marked by the 
second localization process. 
This observation is consistent with the findings of a 
recent numerical study of dipolar colloidal gels \cite{miller}, although in that
study the role of bonds and network junctions has not been quantitatively  
investigated. We suggest that this type of glassy dynamics induced by the 
persistent network might be a distinctive feature of colloidal 
gelation. 

\begin{figure}
\begin{center}
\includegraphics[width=1.0\linewidth]{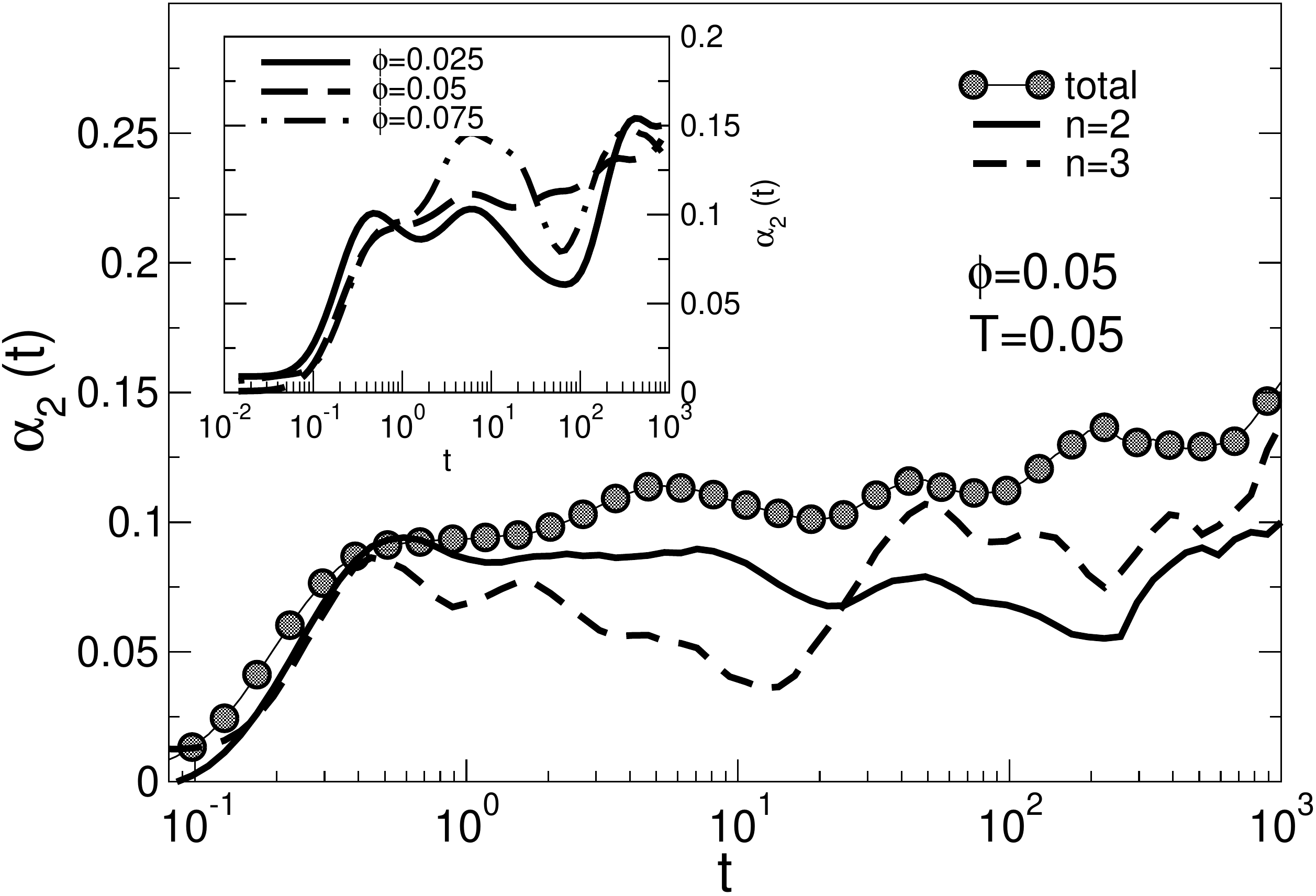}
\caption{ Main frame: Non-Gaussian parameter $\alpha_2(t)$ as a function of time
for $\phi=0.05$ at $T=0.05$.
Inset: $\alpha_{2}(t)$ at $T=0.05$ and volume fractions $\phi=0.025$ 
(full line), $\phi=0.05$ (dashed line) and $\phi=0.075$ 
(dash-dotted line).}
\label{alpha2_structure}
\end{center}
\end{figure}

It is also useful to distinguish the contribution to $\alpha_2(t)$ 
coming from different part of the network structure as we do in 
Fig.~\ref{alpha2_structure}: In the main frame we show $\alpha_{2}(t)$, 
as calculated from all the 
particles at $\phi=0.05$ and $T=0.05$, and compare it to the 
contribution coming from chains particles ($n=2$) and nodes ($n=3$). 
In spite of the fact that the different parts of the structure have been shown 
to give very different contributions to the dynamics, the data indicates that, 
at this low temperature, the $\alpha_{2}(t)$ for the different connectivities 
are 
quite similar. This finding is fully consistent with a recent 
experimental analysis of dynamical heterogeneities in colloidal gels 
of Ref.~\cite{solomon}, where it has been proposed as an indication that 
dynamical heterogeneity in presence of a persistent network 
does not originate from the heterogeneous structure 
but is instead due to the presence of cooperative processes 
as in dense glassy systems \cite{dh}. 
Here we have been able to show that the mechanism producing this glassy regime
is the formation of the persistent, open network. In the following 
sections we will better elucidate the nature of the cooperative processes.
For this we use $F(q,t)$ to investigate the time 
correlations of particle motions over different length
scales in order to understand the connection between 
structural and dynamical heterogeneities in this system.  
\subsection{Incoherent scattering function}
\label{dynamicsc}
\begin{figure}
\begin{center}
\includegraphics[width=1.0\linewidth]{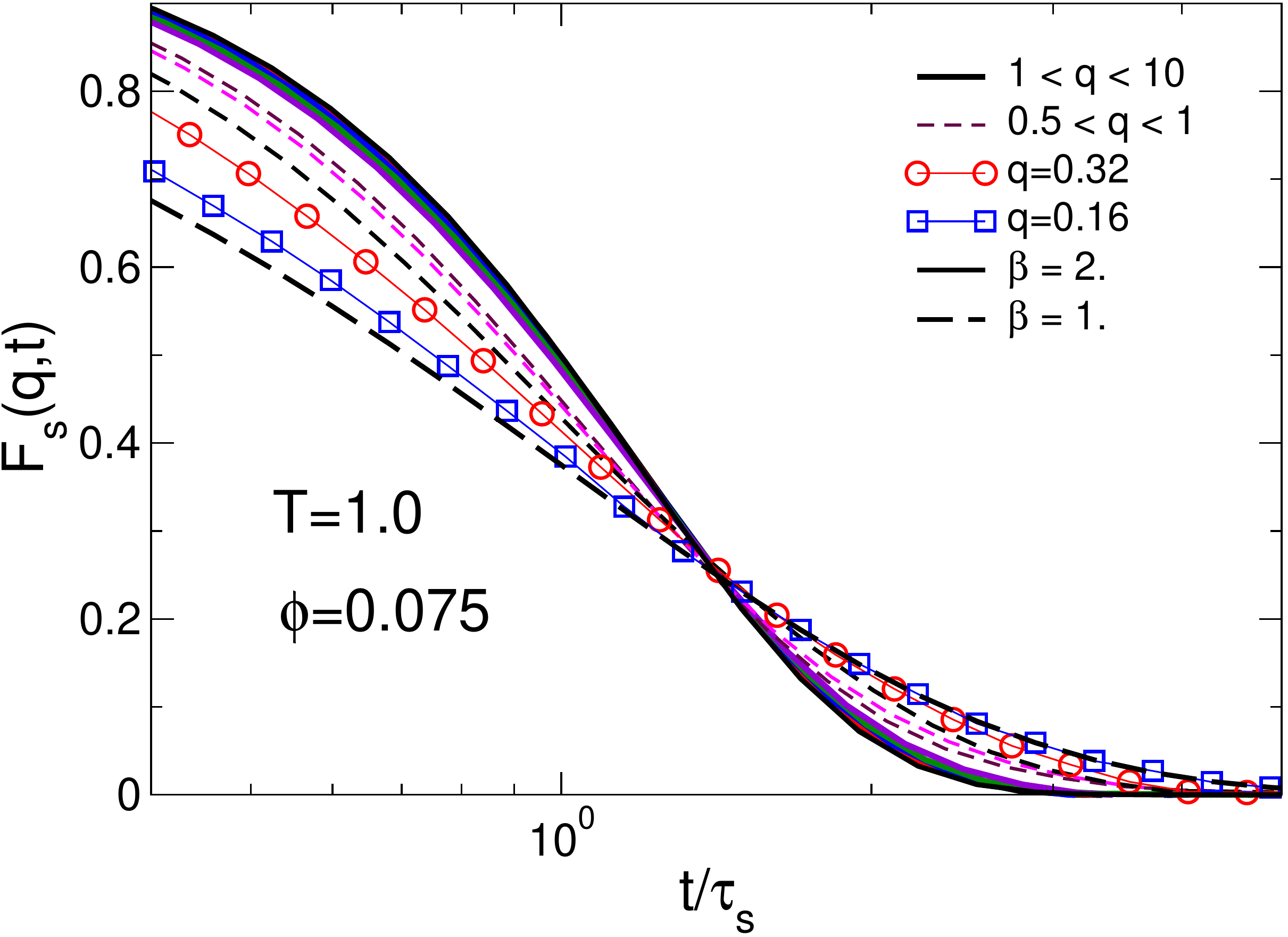}
\caption{
$F_{s}(q,t)$ as a function of time at $T=1.0$ and 
$\phi=0.075$ for wave vectors $q=10.0$, $7.0$, $5.0$, $3.0$, $2.0$, $1.0$, 
$0.88$, $0.54$, $0.32$, $0.16$. The lines correspond to 
fitted decays $\propto \exp(-(t/\tau(q))^\beta)$.
}
\label{fig1_fsq}
\end{center}
\end{figure}

\begin{figure}
\begin{center}
\includegraphics[width=1.0\linewidth]{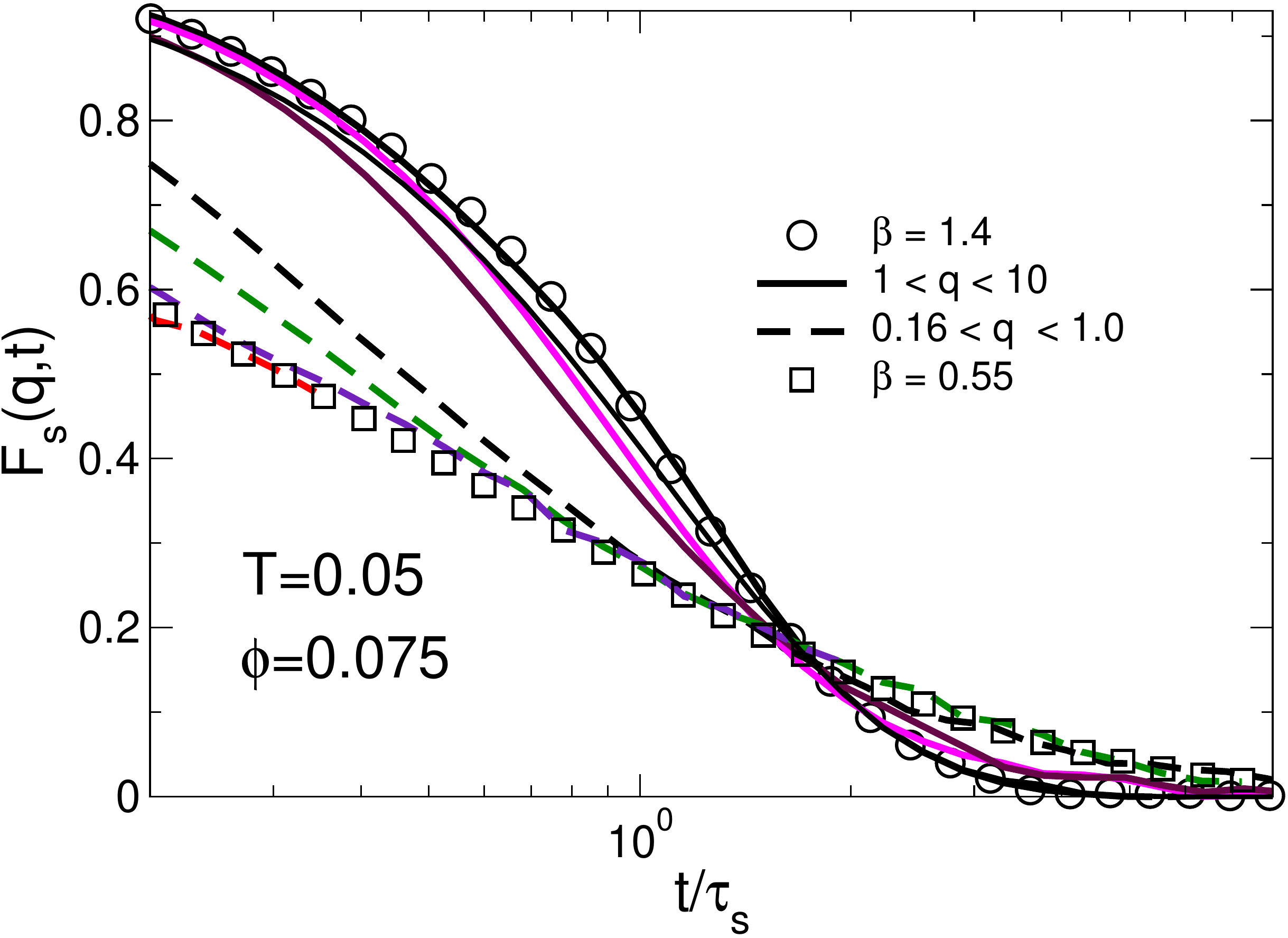}
\caption{
$F_{s}(q,t)$ as a function of time at $T=0.05$ and $\phi=0.075$ for 
wave vectors $q=10.0$, $7.0$, $5.0$, $3.0$, $2.0$, $1.0$,
$0.88$, $0.54$, $0.32$, $0.16$. The open symbols correspond to decays $\propto 
\exp(-(t/\tau(q))^\beta)$.   
}
\label{fig2_fsq}
\vspace{0.5cm}
\end{center}
\end{figure}

In Figs.~\ref{fig1_fsq} and \ref{fig2_fsq} the incoherent scattering function 
$F_{s}(q,t)$, as defined in Eq.~(\ref{self}), is plotted as a function of the 
rescaled time $t/\tau_{s}(q)$ for different wave vectors at $\phi=0.075$, and
at $T=1.0$ and $T=0.05$ respectively. In agreement with the analysis performed 
in Ref.\cite{delgado_kob_prl} at $\phi=0.05$, 
here we also observe that at high 
temperature for $q > 1.0$ all the curves follow a decay 
$\propto \exp(-(t/\tau(q))^{\beta})$ with $\beta=2$ and for $q \leq 1.0$ they 
cross over to a decay with $\beta=1$. This simply illustrates the crossover from 
the ballistic to the diffusive regime, taking place on a length scale of the 
order of the mean free path and is coherent with the  
information obtained from Figs.~\ref{tscales2}, \ref{msd1}, and \ref{msd2}.
At low temperature instead, Fig.~\ref{fig2_fsq} shows a sharper 
crossover, as a function of $q$, from a decay with $\beta \simeq 1.4$ for 
$q > 1.0$ to a decay with 
$\beta \simeq 0.55$ for $q \le 1.0$ (see also 
Ref.~\cite{delgado_kob_prl,kob_sastry_09}). 
At this $T$, finite clusters and free 
particles are rare and the main contribution to the displacement
of the particles comes from the particles of the network which are 
not directly connected to the nodes.
This indicates that, for wave vectors $q$ corresponding to a few
inter-particle distances and smaller, the motion of the branches of
the percolating cluster is the main relaxation mechanism.
For intermediate and large length scales the system shows a relaxation 
dynamics similar to the one found in dense glass-forming liquids, i.e. 
the presence of the disordered structure on this length scales leads to 
heterogeneous dynamics characterized by a stretched exponential decay 
of time correlations~\cite{delgado_kob_prl,delgado_kob_jnnfm}.

\begin{figure}
\begin{center}
\includegraphics[width=1.0\linewidth]{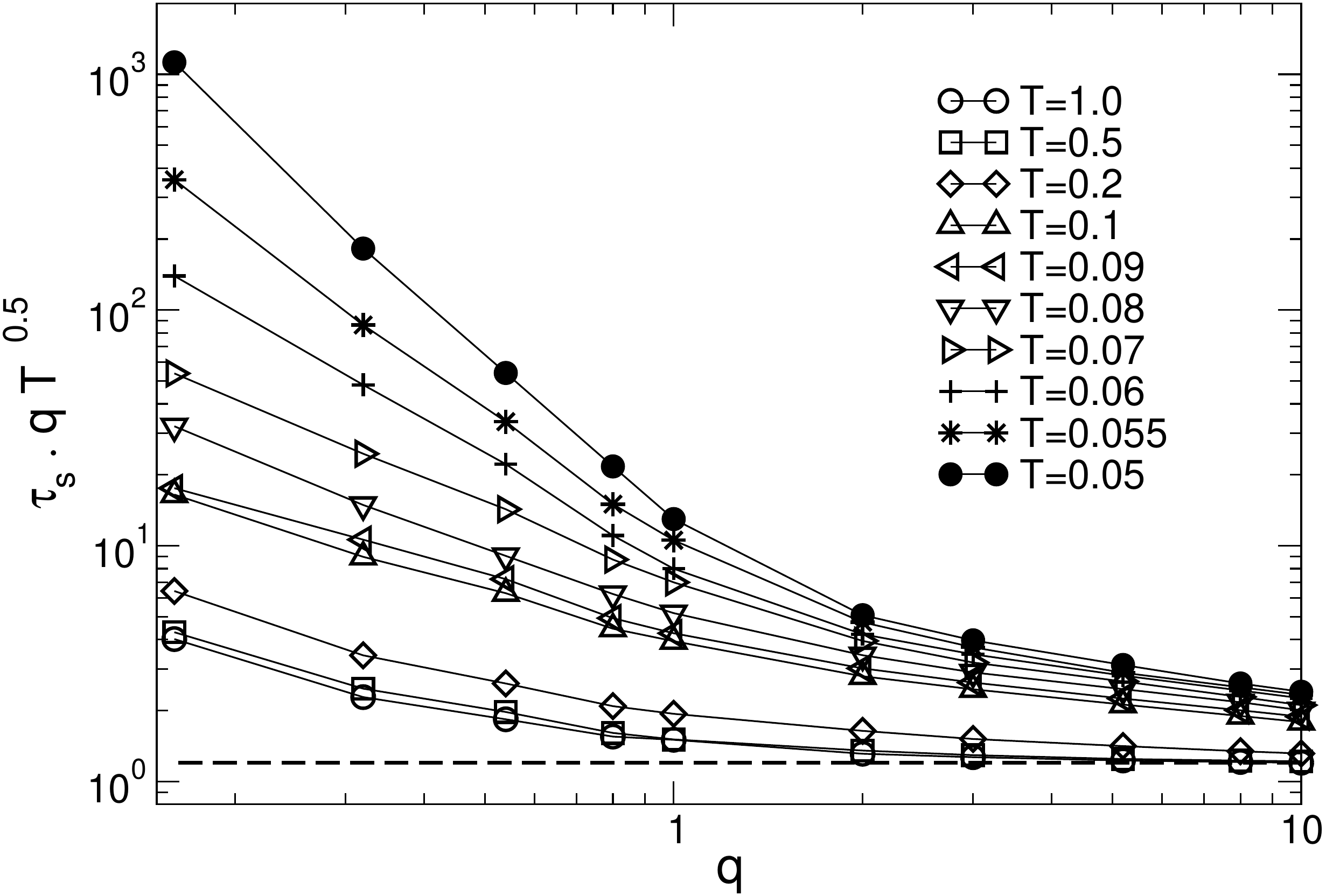}
\caption{
$\tau_{s}(q) \cdot q T^{0.5} $ as a function of the wave vector $q$ 
at different temperatures and $\phi=0.075$ to illustrate the length 
scale dependence of the relaxation processes, at high and 
low $T$. The horizontal line corresponds to 
$\tau_{s}q\sqrt(T)=\sqrt{\pi m/2T}$.}
\label{tscales3}
\end{center}
\end{figure}

In Fig.~\ref{tscales3} we plot $\tau_{s} \cdot q \sqrt{T}$ as a function of $q$ at 
$\phi=0.075$ for different temperatures. 
The horizontal line in the figure, which approximates well the data for high 
$T$ and high $q$, corresponds to $\tau_s= (\sqrt{\pi m /2T})/q$.
At the lowest $T$, the data for large $q$ appears to follow a
different {\it nearly}-ballistic regime. 
For $T\leq0.05$ more than $97\%$ of the particles belong to one percolating
cluster (see Fig.~5), therefore this fast regime is not really
ballistic and is instead due to the motion of the branches of the gel
network \cite{footnote}.
The length scale $q \simeq 1.0$ marks the crossover to a different
dynamic regime. Our data shows that at $T=0.05$ the relaxation time extracted 
from $F_{s}(q,t)$ displays a strong $q-$dependence, which
resembles the one found in dense glass-forming liquids. However, here this 
dependence is observed
at wave-vectors that correspond not to an inter-particle distance, but to
the mesh size of the network.

\begin{figure}
\begin{center}
\includegraphics[width=1.0\linewidth]{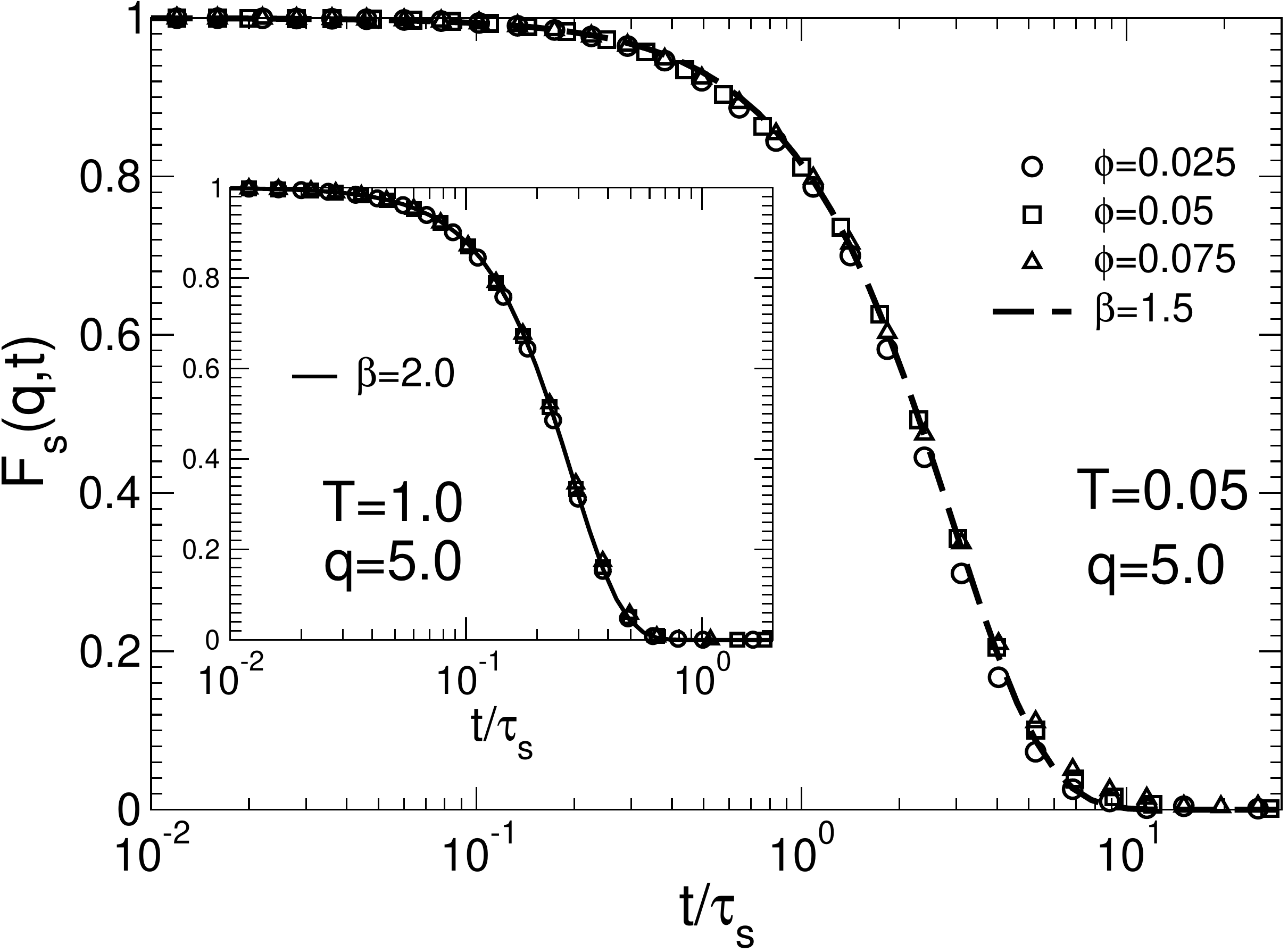}
\caption{
$F_{s}(q,t)$ as a function of rescaled time at $T=1.$ (inset) and 
$T=0.05$ (main frame) for $q=5.0$ and different volume fractions. The lines 
correspond to decays $\propto \exp(-(t/\tau(q))^\beta)$.
}
\label{fig4_fsq}
\end{center}
\end{figure}

\begin{figure}
\begin{center}
\includegraphics[width=1.0\linewidth]{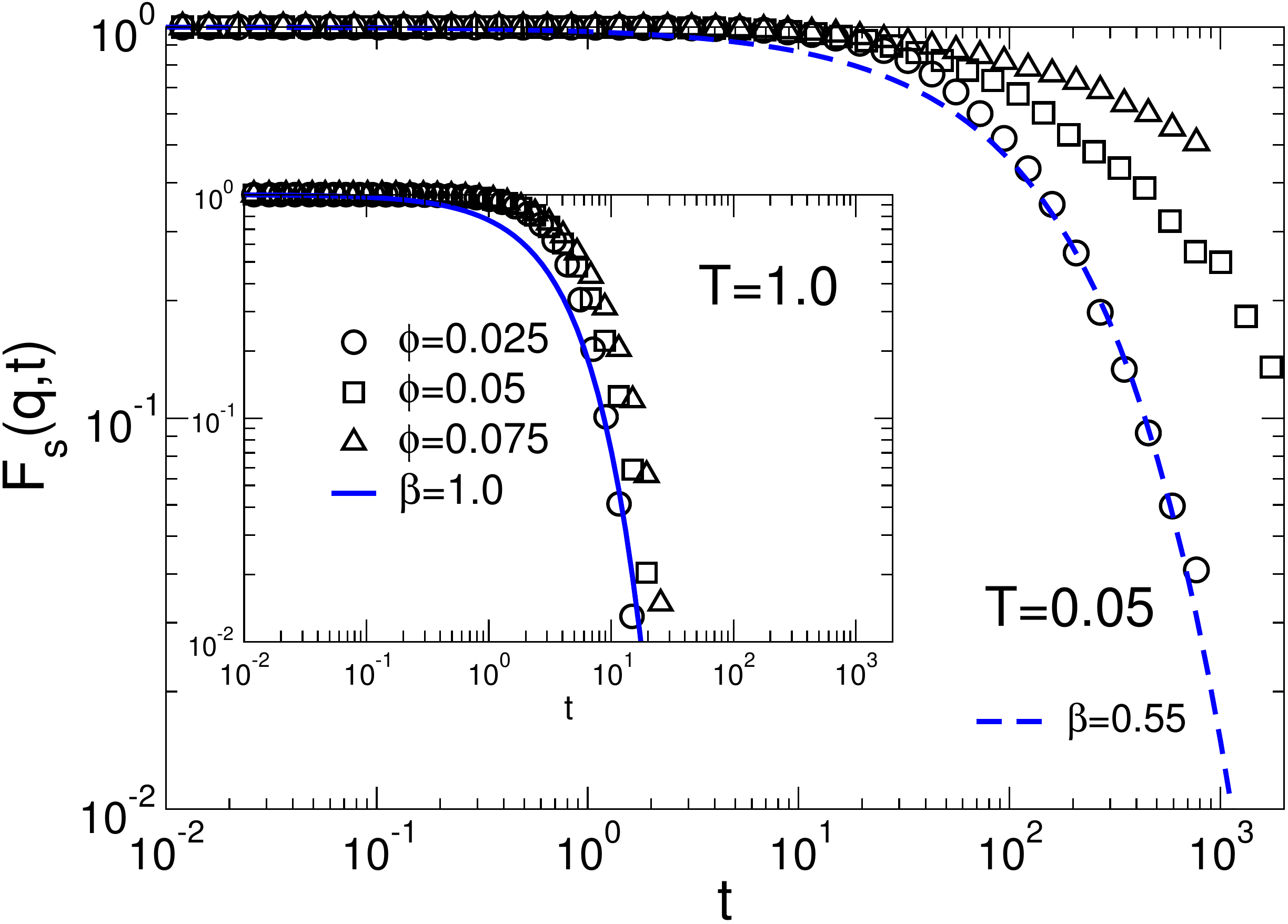}
\caption{
$F_{s}(q,t)$ as a function of time at $T=1.0$ (inset) and
$T=0.05$ (main frame) for $q=0.3$ and different volume fractions.
}
\label{fig3_fsq}
\end{center}
\end{figure}

The complex structure of the network and its strong heterogeneity
has also a detectable effect on the $\phi$-dependence of $F_{s}(q,t)$
at different wave-vectors. In Fig.~\ref{fig4_fsq}, 
$F_{s}(q,t)$ is plotted as a function of time for the different volume 
fractions $\phi= 0.025$, $0.05$, and $0.075$ at high (inset) and low (main frame) 
temperatures for $q=5.0$. 
The plots indicate the change in the time decay from high to low $T$, but the 
$\phi$-dependence is very weak. Although this 
is at the end not surprising in the low volume fraction regime chosen, 
Fig.~\ref{fig3_fsq} clearly shows that this is not obvious. There, $F_{s}(q,t)$
is plotted at high (inset) and low (main frame) temperatures for $q=0.3$. 
On these length scales, one can detect not only the change in the time decay
from high to low $T$, but also a change towards a rather strong 
$\phi$-dependence at low $T$ as compared to high $T$. 
Thus, the emerging picture is
basically the dynamical analogue of Fig.~\ref{fig4}, indicating that,
at small length scales, i.e. up to distances of the order of $3$-$4$ particle
diameters, the structural heterogeneity is completely dominated by the
interaction potential and therefore weakly dependent on $\phi$: this
corresponds to a weakly $\phi$-dependent relaxation dynamics. At larger
length scales instead, beyond the mesh size of the network, even relatively
small changes of $\phi$ strongly affect spatial correlations and
the structure of the network, producing a strongly $\phi$-dependent
relaxation dynamics.
\subsection{Bond and node time correlations}
\label{dynamicsd}
\begin{figure}
\begin{center}
\includegraphics[width=1.0\linewidth]{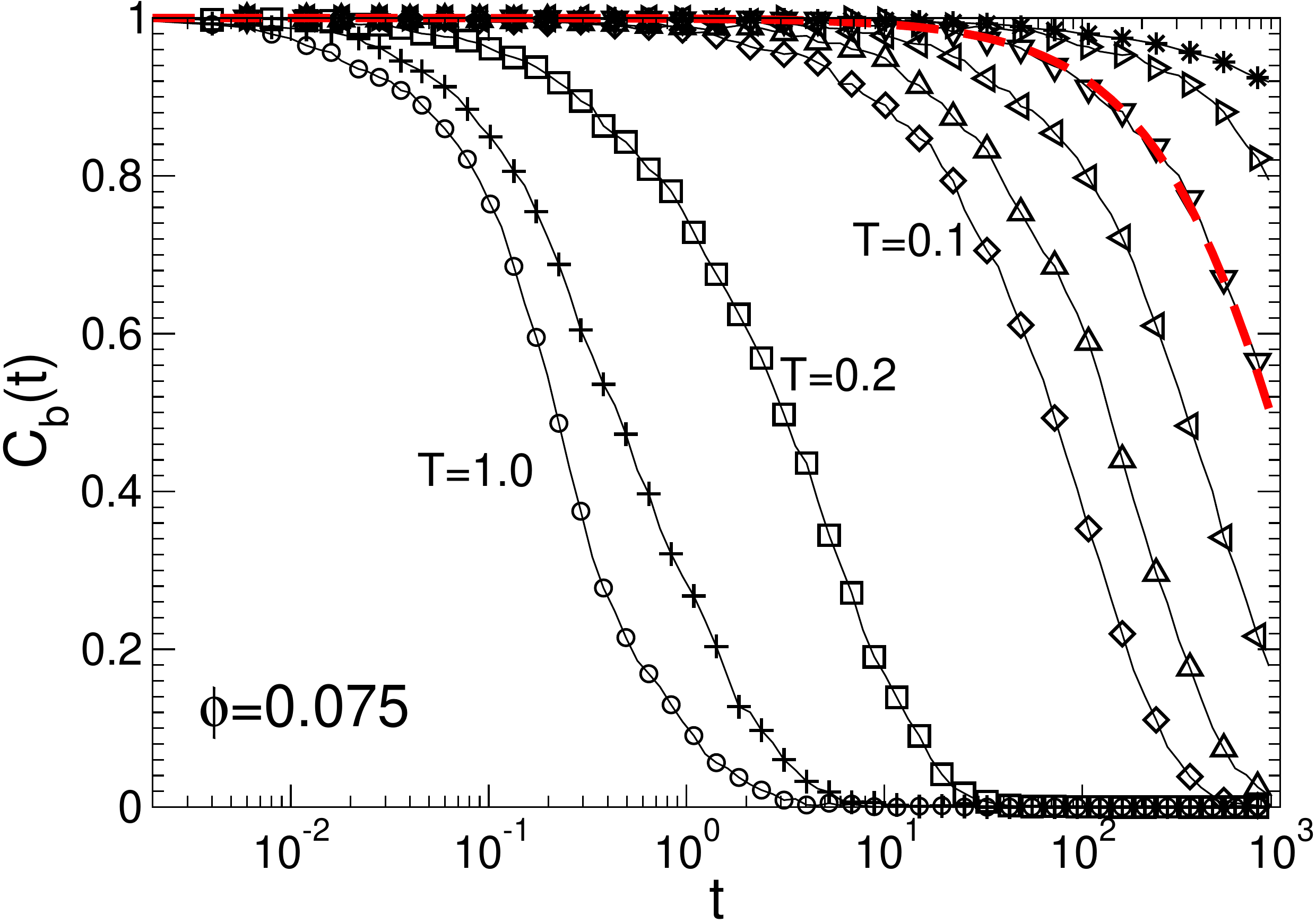}
\caption{The bond time correlation function $C_{b}(t)$ as a function of time,
at volume fraction $\phi=0.075$. $T=1.0$, $0.5$, $0.2$, $0.1$, $0.09$, $0.08$, 
$0.07$, $0.06$, and $0.05$ from left to right. 
The long time decay follows an exponential 
law at all temperatures.
}
\label{bond1}
\end{center}
\end{figure}

\begin{figure}
\begin{center}
\includegraphics[width=1.0\linewidth]{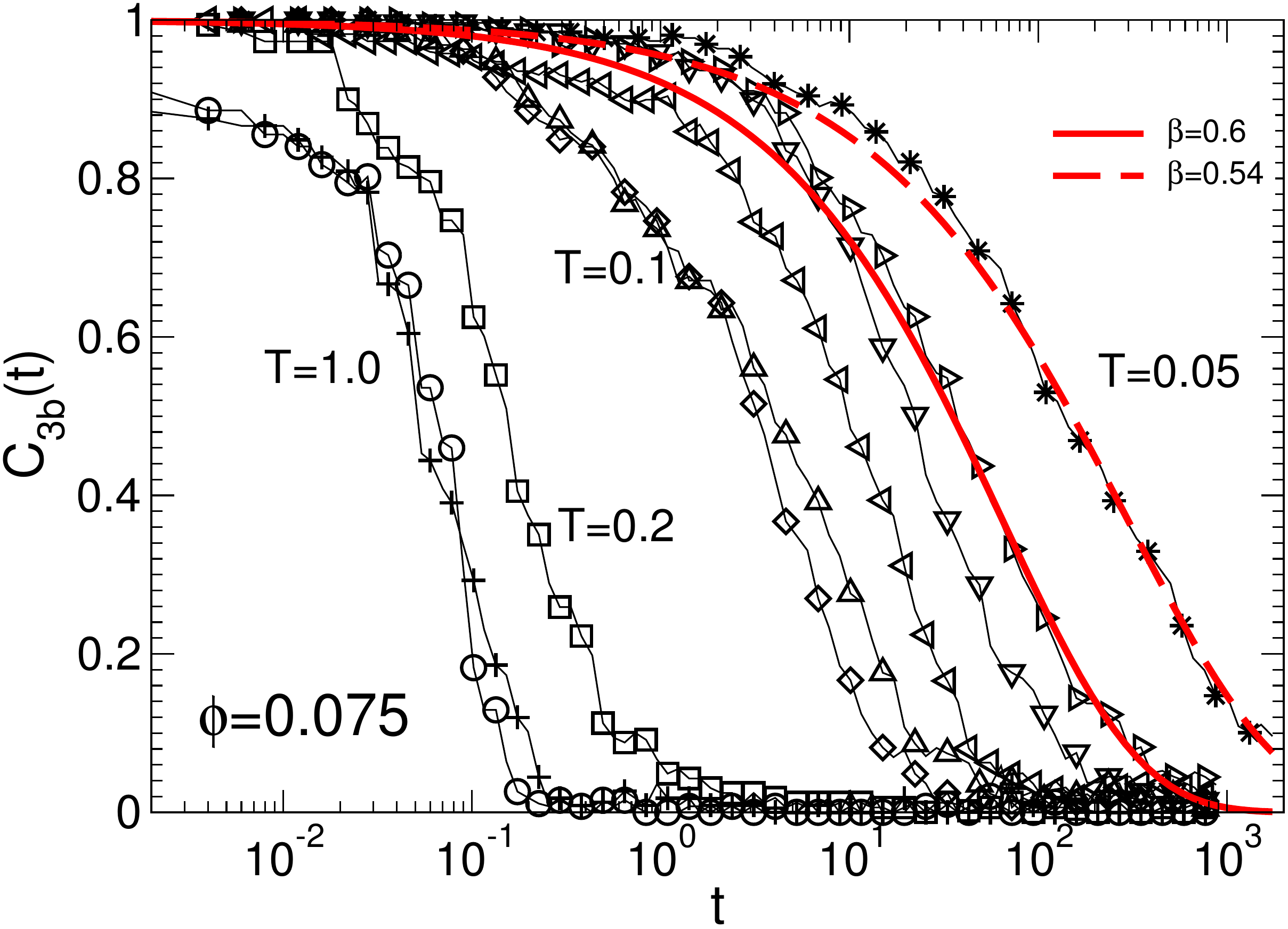}
\caption{
The time correlation function $C_{3b}(t)$ as a function 
of time, at volume fraction $\phi=0.075$. $T=1.0$, $0.5$, $0.2$, $0.1$, 
$0.09$, $0.08$, $0.07$, $0.06$, and $0.05$ from left to right. 
At the lowest temperatures, where it is the time correlation function 
of the network nodes, $C_{3b}(t)$ displays a strongly stretched decay 
($\beta=0.54$).
}
\label{bond2}
\end{center}
\end{figure}

The complex relaxation behavior just described has also its correspondence in 
the decay of time correlation functions for bonds and nodes of the network, as 
elucidated in Figs.~\ref{bond1} and \ref{bond2}. Here, $C_{b}(t)$ and 
$C_{3b}(t)$, as defined respectively in Eqs.~(\ref{bond}) and (\ref{bond3}),
are plotted as a function of time at volume fraction $\phi=0.075$. 
The long time decay of $C_{b}(t)$ is well described by a single exponential 
law at all temperatures, in agreement with the Arrhenius dependence of 
the bond lifetime shown in Fig.~\ref{tscales2}. 
The activation energy related to bond breaking does not significantly 
depend on the volume fraction, as expected.  
As already noticed in Ref.~\cite{delgado_kob_jnnfm}, we can actually distinguish
two different regimes, one at high temperatures ($T > 0.1$) and a second one at low
temperature ($T \leq 0.1$). At high temperatures the 
particle collisions promote uncorrelated bond breaking or formation, 
leading to a short time decay of bond correlation and hence to a quick 
decay of correlations.
At low temperatures instead, the role of particle collisions decreases and 
the energy activation process becomes
the only relevant process in the decay of bond correlation. 
In Fig.~\ref{bond2}, the time correlation function $C_{3b}(t)$ for particles 
with coordination number $n=3$ is plotted as a function of time.
At high temperatures ($T > 0.1$) particles of connectivity $3$ are 
rare (see 
Fig.~\ref{coord}), and therefore the statistics is rather poor. 
At temperatures $T\geq 0.07$, the long time decay of time
correlation functions $C_{3b}(t)$ follows a simple exponential law, with a
characteristic relaxation time $\tau_{3b}(\phi,T)$ increasing with 
decreasing $T$. This is coherent with the Arrhenius dependence of $\tau_{3b}$
in Fig.\ref{tscales2}. In contrast to this, at temperatures at which a 
persistent 
spanning network is present in the system, i.e. $T \le 0.06$,
the decay of $C_{3b}(t)$ becomes stretched, with a
stretching exponent $\beta$ which decreases with $T$ ($\beta \simeq 0.54$ at
$T=0.05$). This indicates that, once the network is formed and it is
sufficiently persistent, the process of breaking and formation of the nodes is 
associated not only to the overcoming of an activation energy but also to a
heterogeneous and cooperative dynamic process \cite{delgado_kob_05}.

We have further investigated the type of cooperative processes in which the 
nodes might be involved, in particular by measuring the occurrence of events 
at which the breaking of one node makes that one of its neighbors of 
connectivity $2$ becomes itself a node. This would make the nodes to slide 
along the chains. We have found that in the network regime at least 
$25\%$ of the events corresponds to this situation. 
This behavior points out a further similarity, in spite of the very different type of interactions, 
 between the system studied here and 
the dipolar colloidal gels of Ref.~\cite{miller}, in which one could easily 
expect these events to occur. 
These results elucidate well the complex interplay between structure and 
dynamics which is fundamental in these systems: Although the slow 
dynamics is due to the bond lifetime becoming 
sufficiently long (see Fig.\ref{tscales2}),
it is only the combination of sufficiently persistent bonds 
together with the branching of the aggregates, 
i.e. the formation of a stress bearing network, 
which eventually determines 
the arising of large scale cooperative relaxation processes at these low volume
fractions.

\subsection{Summary and discussion of the dynamical properties}
\label{dynamicse}
We can now put together all the elements obtained from the different quantities
into a unique coherent picture. The onset of the aggregation, 
signaled for example by the change of shape of the cluster size distribution 
around $T=0.1$ in Fig.~5, corresponds to the onset of a dynamical 
regime at which relaxation processes are dominated by bond-breaking as opposed to
particle collisions (Fig.~\ref{tscales2}). This dynamical regime
may resemble, to some extent, the onset of caging in dense 
systems: The MSD, 
signals a localization process over similar length scales 
(Figs.~\ref{msd1} and \ref{msd2}) accompanied by a growing degree of 
heterogeneity in the 
distribution of particle displacements (Figs.~\ref{alpha2_1} and 
\ref{alpha2_structure}), as compared to a Gaussian one. 
In spite of these similarities, we have shown that here 
this regime does not imply 
structural arrest. The heterogeneity of the dynamics is clearly 
arising from the wide distribution of the sizes of the aggregates
~\cite{delgado_kob_prl,pablo,tiziana_prl,tiziana_pre}. 
Upon further decreasing temperature, 
the aggregation process leads to the formation of an interconnected network. 
Once that the network is 
persistent enough, i.e. the node lifetime starts to be comparable to the
longest relaxation times in the system (Fig.~\ref{tscales2}), a 
new, slow dynamic regime sets in. This has certainly the hallmark 
of gelation, due to the formation of the locally rigid, persistent network,
and it is also indicated by the strong localization process in the MSD, 
strikingly dominated by the network nodes (inset of Fig.~\ref{msd1}),
over length scales of the order of the mesh size of the network.
This second dynamical regime is associated to a qualitative change in
dynamical heterogeneity (Fig.~\ref{alpha2_1}). 
The contribution arising from the first localization process
(caging) becomes negligible: there is a new significant non-Gaussian 
contribution to the distribution of particle displacements 
arising from the second localization process (see the curve at $T=0.07$ in 
Fig.~\ref{alpha2_1}) and moving towards longer and longer time scales. 
The heterogeneity of this slow dynamics has no straightforward relation 
to the structure (Fig.~\ref{alpha2_structure}), therefore suggesting 
that it signals instead the presence of new cooperative processes. 
The study of the relaxation dynamics over different length scales 
elucidates well that this second dynamical regime is characterized by 
the coexistence of very different relaxation processes over different length 
scales (Figs.~\ref{fig1_fsq}-\ref{fig3_fsq}): fast motion of pieces 
(chains and dangling ends) of the gel at small distances and 
slow, stretched exponential processes related to the network rearrangements
at length scales larger than the network mesh size. This clearly indicates 
the network origin of this dynamical regime, which, being characterized by 
slower and slower, complex relaxation, points to structural arrest.
Our comparative analysis indicates therefore that this complex dynamics
has glassy features and strong similarities to the one of dense systems, 
with the difference that the strong coupling in particle motion (which originates 
the structural arrest) is not induced by the crowding but
by the presence of the persistent network. 
The fact that the onset of very slow, stretched exponential relaxations
only takes place at sufficiently low wave vectors is a consequence of that.
Finally, by looking to bond and nodes relaxation we have found 
that the cooperative processes underlying this network induced glassy dynamics
is intimately related to the network nodes (Fig.~\ref{bond2}).\\

\section{Conclusions}
\label{conclu}
We have discussed the behavior of a model for colloidal gels 
based on the presence of directional effective interactions. 
In this model, the aggregation leads to an open persistent network structure
without imposing a fixed connectivity to the gel units. 
With these premises, our study gives new insights into the 
physics of colloidal gelation. From the structural point of view, 
we have shown that the aggregation process takes place via a random 
percolation mechanism, but once a percolating structure is formed, 
it rapidly evolves towards a persistent, fully connected open network.
Also in this case, gelation seems associated to the presence of a pre-peak
in the static structure factor, here 
directly related to the network structure.
Our comparative analysis of structure and dynamics indicates that 
the persistent network introduces slow, cooperative processes 
intimately related to the network nodes and sets in a peculiar 
kind of glassy dynamics. 
This scenario is also consistent with the results of 
a recent numerical study of diluted dipolar colloidal gels~\cite{miller}. 
We think that this points to the formation of a persistent network 
as the mechanism responsible for the onset of the glassy dynamics in colloidal 
gels and explains the close connection between gelation and glassy structural 
arrest typically observed in these systems. We therefore suggest 
that this scenario may in fact be relevant to the physics of colloidal gels 
on a more general ground.  

{\it Acknowledgments}: The authors would like to thank L. Cipelletti
for many fruitful discussion, 
and M.E. Cates for interesting suggestions. Part of this work was  
supported by the Marie Curie Fellowship MCFI-2002-00573 and by ANR-TSANET.


\begin{thebibliography}{99}
\bibitem{footnote}
As a side remark, we would like to comment that the microscopic Newtonian
dynamics we use is certainly not the most appropriate to the damped
microscopic dynamics of a real system, due to the presence of the solvent.
From this point view, the use of Brownian dynamics would be more appropriate.
On the other hand, the choice of the Newtonian dynamics has been useful
here because it allows for a clearer distinction of the two different
relaxation regimes and has therefore simplified our analysis.
\end{thebibliography}
\end{document}